\documentclass[draft]{agujournal2019}
\usepackage{ragged2e}
\usepackage{url} %this package should fix any errors with URLs in refs.
\usepackage{lineno}
\usepackage[inline]{trackchanges} %for better track changes. finalnew option will compile document with changes incorporated.
\usepackage{soul}
\usepackage{comment}
\usepackage{amsmath}
\usepackage{amsfonts}
\usepackage{amssymb}
\usepackage[caption=false]{subfig} %[caption=false] was added to bypass errors shown in caption.
\usepackage{adjustbox} %to rotate table
\usepackage[graphicx]{realboxes} %to rotate table

\draftfalse
\journalname{Space Weather}

\begin{document}
\justifying

%% ------------------------------------------------------------------------ %%
%  Title
%
% (A title should be specific, informative, and brief. Use
% abbreviations only if they are defined in the abstract. Titles that
% start with general keywords then specific terms are optimized in
% searches)
%
%% ------------------------------------------------------------------------ %%

% Example: \title{This is a test title}

\title{Employing the coupled EUHFORIA-OpenGGCM model to predict CME geoeffectiveness}

%% ------------------------------------------------------------------------ %%
%
%  AUTHORS AND AFFILIATIONS
%
%% ------------------------------------------------------------------------ %%

% Authors are individuals who have significantly contributed to the
% research and preparation of the article. Group authors are allowed, if
% each author in the group is separately identified in an appendix.)

% List authors by first name or initial followed by last name and
% separated by commas. Use \affil{} to number affiliations, and
% \thanks{} for author notes.
% Additional author notes should be indicated with \thanks{} (for
% example, for current addresses).

% Example: \authors{A. B. Author\affil{1}\thanks{Current address, Antartica}, B. C. Author\affil{2,3}, and D. E.
% Author\affil{3,4}\thanks{Also funded by Monsanto.}}
%[0000-0002-4269-056X]
\authors{Anwesha Maharana\affil{1,2}, W. Douglas Cramer\affil{3}, Evangelia Samara\affil{4}, Camilla Scolini\affil{2,3}, Joachim Raeder\affil{3}, and Stefaan Poedts\affil{1,5}}

\affiliation{1}{Centre for mathematical Plasma Astrophysics (CmPA)/Dept.\ of Mathematics, KU Leuven, Celestijnenlaan 200 B, 3001 Leuven, Belgium}
\affiliation{2}{Solar–Terrestrial Centre of Excellence—SIDC, Royal Observatory of Belgium, 1180 Brussels, Belgium}
\affiliation{3}{Institute for the Study of Earth, Oceans and Space, University of New Hampshire, 03824 Durham, NH, USA}
\affiliation{4}{NASA Goddard Space Flight Center, Greenbelt, MD, United States}
\affiliation{5}{Institute of Physics, University of Maria Curie-Sk{\l}odowska, ul.\ Radziszewskiego 10, 20-031 Lublin, Poland}

%\affiliation{=number=}{=Affiliation Address=}
%(repeat as many times as is necessary)

%% Corresponding Author:
% Corresponding author mailing address and e-mail address:

% (include name and email addresses of the corresponding author.  More
% than one corresponding author is allowed in this LaTeX file and for
% publication; but only one corresponding author is allowed in our
% editorial system.)

% Example: \correspondingauthor{First and Last Name}{email@address.edu}

\correspondingauthor{Anwesha Maharana}{anwesha.maharana@kuleuven.be}

%% Keypoints, final entry on title page.

%  List up to three key points (at least one is required)
%  Key Points summarize the main points and conclusions of the article
%  Each must be 140 characters or fewer with no special characters or punctuation and must be complete sentences

% Example:
% \begin{keypoints}
% \item	List up to three key points (at least one is required)
% \item	Key Points summarize the main points and conclusions of the article
% \item	Each must be 140 characters or fewer with no special characters or punctuation and must be complete sentences
% \end{keypoints}
\begin{keypoints}
    \item Coupling of EUHFORIA with OpenGGCM self-consistently provides physics-based predictions of the geo-effectiveness of CME impacts.
    \item Geomagnetic indices can be predicted 1-2 days in advance with EUHFORIA, as opposed to the much shorter predictions with real-time L1 data.
    \item We employ an advanced metric, Dynamic Time Warping (DTW), to account for time shifts, and to assess the model performance in predicting Dst.
\end{keypoints}

%% ------------------------------------------------------------------------ %%
%
%  ABSTRACT and PLAIN LANGUAGE SUMMARY
%
% A good Abstract will begin with a short description of the problem
% being addressed, briefly describe the new data or analyses, then
% briefly states the main conclusion(s) and how they are supported and
% uncertainties.

% The Plain Language Summary should be written for a broad audience,
% including journalists and the science-interested public, that will not have 
% a background in your field.
%
% A Plain Language Summary is required in GRL, JGR: Planets, JGR: Biogeosciences,
% JGR: Oceans, G-Cubed, Reviews of Geophysics, and JAMES.
% see http://sharingscience.agu.org/creating-plain-language-summary/)
%
%% ------------------------------------------------------------------------ %%

%% \begin{abstract} starts the second page

\begin{abstract}
EUropean Heliospheric FORecasting Information Asset (EUHFORIA) is a physics-based data-driven {solar} wind and CME propagation model designed for space weather forecasting and event analysis investigations. Although EUHFORIA can predict the solar wind plasma and magnetic field properties at Earth, it is not equipped to quantify the geo-effectiveness of the solar transients in terms of geomagnetic indices like the disturbance storm time (Dst) index and the auroral indices that quantify the impact of the magnetized plasma encounters on Earth’s magnetosphere. Therefore, we couple EUHFORIA with the Open Geospace General Circulation Model (OpenGGCM), a magnetohydrodynamic model of the response of Earth’s magnetosphere, ionosphere, and thermosphere to transient solar wind characteristics. In this coupling, OpenGGCM is driven by the solar wind and interplanetary magnetic field obtained from EUHFORIA simulations to produce the magnetospheric and ionospheric response to the CMEs. This coupling is validated with two observed geo-effective CME events driven with the spheromak flux-rope CME model. We compare these simulation results with the indices obtained from OpenGGCM simulations driven by the measured solar wind data from spacecraft. We further employ the dynamic time warping (DTW) technique to assess the model performance in predicting Dst. The main highlight of this study is to use EUHFORIA simulated time series to predict the Dst and auroral indices 1 to 2 days in advance, as compared to using the observed solar wind data at L1, which only provides predictions 1-2 hours before the actual impact. 
\end{abstract}

\section*{Plain Language Summary}
Coronal mass ejections (CMEs) are gigantic eruptions of magnetized and charged matter (plasma) from the Sun that can propagate all the way to Earth and interact with Earth's magnetic shield, the magnetosphere. They can cause major disruptions in space- and ground-based technologies, and in turn, affect our socio-economic lives on Earth. Hence, we predict the arrival time of the CMEs and quantify the impact of such geomagnetic storms. %, is a crucial aspect of space weather (the conditions of the near-Earth space environment in the presence of electromagnetic radiation and charged particles) forecasting. 
Even with the 24x7 monitoring of the Sun from Earth's perspective, it is not possible to accurately predict the arrival times of Earth-directed CMEs, and most importantly, assess severity of the associated storms. % due to the challenges in reconstructing and understanding their morphology from just the line-of-sight observations. %Serendipitous observing by multiple spacecraft is not reliable for forecasting purposes. 
Therefore, we perform 3D physics-based modeling with the EUHFORIA model to track the CMEs from the Sun to Earth. EUHFORIA provides the plasma and magnetic field conditions near Earth. However, with just EUHFORIA output we cannot quantify the geomagnetic impact of the CMEs. Therefore, EUHFORIA output is used as input to OpenGGCM, a 3D physics-based model that simulates the response of magnetosphere, ionosphere, and thermosphere to CME impact. %OpenGGCM not only models the interaction of the CMEs with the magnetosphere but also trickles down the calculations into the ionosphere and thermosphere to give an overall impact on Earth's upper atmosphere. 
%In practice, geomagnetic indices like the Dst and auroral indices are obtained 20-90~minutes before the actual CME impact is observed. 
%Current operational models provide only 20-90 minutes lead time for predicting geomagnetic indices like the Dst and auroral indices. Such a short prediction window limits our mitigation actions. 
With the joint EUHFORIA and OpenGGCM simulations, geomagnetic indices can be predicted 1-2 days in advance, thus making this a promising methodology for CME-induced space weather forecasting. %In this work, we validate the coupling of EUHFORIA and OpenGGCM with two observed geomagnetic storms related to CMEs, and assess its prediction capability.

%% ------------------------------------------------------------------------ %%
%
%  TEXT
%
%% ------------------------------------------------------------------------ %%
%%%%%%%%%%%%%%%%%%%%%%%%%%%%%%%%%%%%%%%%%%%%%%%%%%%%%%%%%%%%%
%
%-------------------------Section---------------------------%
%
%%%%%%%%%%%%%%%%%%%%%%%%%%%%%%%%%%%%%%%%%%%%%%%%%%%%%%%%%%%%%
\section{Introduction}\label{sec:intro}
The diverse socio-economic and health impacts of solar-terrestrial relations have drawn attention towards geomagnetic storms and space weather forecasting \cite{FREng2007,Oughton2019,Pilipenko2021,Zenchenko2021,Buzulukova2022}. The solar wind causes spatiotemporal variability in the geospace environment, especially the magnetosphere and the ionosphere \cite[ and references therein]{Chapman1931,Akasofu1963,Akasofu2021}. %Although Earth's magnetosphere provides natural protection from the solar phenomena, under certain circumstances the space weather effects are exacerbated upon the arrival of solar transients like coronal mass ejections (CMEs), stream interaction regions (SIRs), and solar energetic particles (SEPs) at Earth. 
Earth's magnetosphere provides natural protection from the solar phenomena. However, a southward interplanetary magnetic field (IMF) caused by solar transients like coronal mass ejections (CMEs) can result in a rapid magnetic reconnection with Earth's northward pointing {magnetic field in the day side of the magnetosphere} causing geomagnetic storms \cite{Chapman1918,Dungey1961,Bothmer2007}. %In the southward reconnection case, the reconnection region stretches across the entire dayside magnetopause as compared to the northward case where the region of anti-parallel field lines is small in the nightside \cite{Bothmer2007}.
 {A geomagnetic storm is characterized by three phases. First is the sudden commencement, when Earth's magnetic field is suddenly compressed, second is the main phase when the magnetic field rapidly decreases, and finally in the recovery phase the magnetic field strength returns back to normal \cite<Chapter 4; >{Bothmer2007}.} Moreover, when the two reconnection sites (the one on the day side and the other one on the night side) are sufficiently separated, different parts of the magnetosphere respond to the storm at different times, leading to intermittent storage and release of energy. This results in a phenomenon called magnetospheric substorm \cite{Akasofu1968}. These auroral/magnetospheric substorms are also linked to the ring current development during the main phase of the storm \cite{Akasofu2020,Alberti2022}. The substorms also interfere with GPS communication and impact electric power systems  \cite[and references therin]{Skone2000,Boteler2001}. 

{The process of using the ground magnetic field observations directly in real-time, to evaluate geomagnetic indices (like the Disturbance storm time (Dst) index, auroral indices or A-indices - AU, AL, AE, and Kp index) is called `nowcasting' or `short-term forecasting'. Nowcasting is important for situational awareness, i.e., for verifying whether the occurrence of power outages or satellite malfunctions is a consequence of a geomagnetic storm or if they are caused {by} {terrestrial or man-made }accidents or activities of other socio-political agencies. However, the geomagnetic indices must be forecasted for mitigation purposes. The standard forecasting methodology involves predicting the indices with the near-Earth near-real-time (NRT) solar wind data, i.e.\ using the solar wind conditions at L1 to compute the geomagnetic indices at Earth \cite{Keesee2020,Smith2021,Smith2022}. With NRT data, the earliest warning of impact at Earth that we can obtain at present times, is from the satellites at L1, unless there is a serendipitous multi-spacecraft lineup before 1~au. This gives us less than two hours \cite<20-90 minutes;>{Wintoft2017} to take measures against these storms which is substantially {insufficient} for prompt and effective mitigation for the space- and ground-based technological infrastructure. The ideal lead time for the best mitigation for the power grid operators is 2-3 days and for the aviation industry is 24 hours \cite<{Joint Research Centre (JRC)} Science for Policy Report, {\url{https://publications.jrc.ec.europa.eu/repository/handle/JRC104231}},>{krausmann2016}. This major limitation of a short forecast time window associated with using NRT observations, motivates us to employ modeled solar wind data (predictions) in computing the geomagnetic indices to increase the forecast time window.} 
 
Various state-of-the-art empirical (e.g., HUXt, \citeA{Barnard2022}; DBM, \citeA{vrsnak2013}, etc), and MHD models (e.g., ENLIL, \citeA{Odstrcil2003}; {MAS, \citeA{Mikic1999}}; SUSANOO, \citeA{Shiota2016}, AWSoM, \citeA{vanderholst2014}, EUHFORIA, \citeA{Pomoell2018};  etc) have been created to study CME propagation and predict their arrival time. Heliospheric models like EUHFORIA provide an additional advantage of employing magnetized CME models \cite{Verbeke2019,Maharana2022} for improving the prediction of magnetic field components (especially the z component of IMF) in addition to plasma properties. The first step in the realistic prediction of CME arrival time at Earth is modeling as accurately as possible the ambient solar wind conditions through which the CMEs traverse. The second step involves constraining the initial CME parameters (geometric and magnetic field) from pre- and post-eruption observations \cite{Scolini2019}. The final step is driving the 3D heliospheric simulations with the modeled ambient solar wind conditions and inserting the CMEs as time-dependent boundary conditions. Ideally, with the best and fullest efforts involving the maximum human and machine resources, the arrival time prediction of a CME event must be accomplished within 1--2 days from the launch of the CME from the Sun. If achieved, the magnetosphere models can be employed to make the geomagnetic index forecast 1--2 days in advance of the actual recording of the geomagnetic storm at Earth. %To contextualize this, the Deep Space Climate Observatory (DSCOVR) satellite located at L1 can provide an early warning of CME arrival only 15-60~minutes before the CME reaches Earth ({\url{https://www.swpc.noaa.gov/phenomena/coronal-mass-ejections}}).
This calculation assumes an average actual CME arrival time of 3--4 days \cite{Iwai2021}. The increased lead time of 1-2 days {is much better in comparison with the above-mentioned 20-90 minutes warning time}, and can potentially help in monitoring and planning mitigation strategies more efficiently. For the prediction of even faster CMEs, further improvements are required in individual models in the chain.\\
%Modeling the propagation of CMEs with EUHFORIA, at the first stage, requires the initial chain of models, i.e., corona and heliosphere. This process can take up to 1--2 days (depending on the computational infrastructure, human resources, and physical complexity) as it involves modeling the solar wind, and observationaly constraining the CME parameters for initializing and running the physics-based heliosphere models. This means that for an average speed CME that takes 3--4 days \cite{Iwai2021} to reach Earth, we can obtain the solar wind predictions 1--2 days in advance through simulations. To contextualize this, the Deep Space Climate Observatory (DSCOVR) satellite located at L1 can provide an early warning of CME arrival 15-60~minutes before the CME reaches Earth ({\url{https://www.swpc.noaa.gov/phenomena/coronal-mass-ejections}}). Hence, this is one of the advantages of replacing real-time data with modeled solar wind predictions. The endpoint of these model chains, i.e., magnetosphere and ionosphere models (especially physics-based) can receive the solar wind data 1--2 days in advance instead of minutes which would enable an earlier warning (in the order of hours) than using real-time solar wind data. The increased lead time of 1--2 days can potentially help in monitoring and planning mitigation strategies more efficiently. {For the prediction of even faster CMEs, further improvements are required in individual models in the chain.}\\

The quantification of the comprehensive impact of the solar wind on the geospace, the connected layers of magnetosphere, ionosphere, and thermosphere (MIT), requires the consideration of interactions with the incoming plasma and magnetic field. %Understanding the global nature of the transport and conversion of energy, mass, and magnetic flux between the solar wind and the geospace is not fully self--consistent with empirical models. 
The complexity of the geospace, because of its multiple coupled components, added to the non-linearity and time dependence of the physical processes, demands a numerical approach for its modeling. %Numerical approaches not only help in testing and improving the theoretical framework of the phenomena but also enhance our capabilities in building a better space weather forecasting infrastructure. 
Three-dimensional magnetosphere models like Open Geospace General Circulation Model  \cite<OpenGGCM, >{Raeder1998}, GUMICS \cite{Janhunen1996}, BATSRUS \cite{Powell1999} -- {implemented as the Space Weather Modeling Framework, \cite<SWMF, >{Toth2012}}, and PPMLR-MHD \cite{Hu2007} etc.\ (see the review by \citeA{Wang2013}) complement single point-based observations by satellites or ground-based instruments to understand the complex physics. In a study assessing different types of magnetospheric models, \citeA{Rastatter2013} reported that the empirical models (analytic or iterative formula or neural network-based algorithm without involving calculations of the intrinsic energy flow in the geospace) perform well in Dst prediction, in general. The magnetosphere models that could match the accuracy of the empirical results are the global MHD models of the magnetosphere that include the inner magnetosphere dynamics, i.e., the ring current model or the stand-alone ring current models with well-defined boundary conditions. OpenGGCM performed average or poorly relative to models like BATSRUS and CMIT in the analysis by \citeA{Rastatter2013} because of the absence of a ring current model. However, the latest version of OpenGGCM (v5.0.ccmc) is coupled to a ring current model, Rice Convection model \cite<RCM;>{Toffoletto2003} capable of computing the kinetic ring current \cite{Cramer2017}. In addition, OpenGGCM is coupled to an ionosphere and thermosphere model, which makes it a global model of the geospace. 

The physics of CME eruption, propagation, and its interaction with the magnetosphere happens at different spatial and temporal scales. Coupling of models from the Sun to Earth \cite{Luhmann2004,Toth2007} has become a popular `trend’ in space weather modeling as it enables faster and more efficient space weather predictions at Community Coordinated Modeling Center ({\url{CCMC, }{https://ccmc.gsfc.nasa.gov/}}) and Virtual Space Weather Modelling Centre ({\url{VSWMC, }{https://esa-vswmc.eu/}}) \cite{Poedts2019}. {Combining models working at different spatiotemporal scales conserves the physics at each stage and also consumes less time than running a single model from Sun to Earth.} The objectives of this study are two-fold: (1) To demonstrate the coupling of 3D MHD models of the heliosphere (EUHFORIA) and magnetosphere (OpenGGCM), and (2) to emphasize the possibility of forecasting geomagnetic indices with a chain of physics-based models instead of nowcasting or forecasting with L1 observations that provide a short forecast time window of around an hour. The paper is organized as follows. In Section~\ref{sec:models}, we introduce the heliospheric model of EUHFORIA and the MIT model of OpenGGCM and demonstrate the coupling between them. Section~\ref{sec:geomagnetic_indices} lists and explains the geomagnetic indices computed using this coupling. The real events used for validating the coupling are detailed in Section~\ref{sec:validation}. In Section~\ref{sec:analysis}, we analyze the capability of this coupling to forecast the geomagnetic indices for the chosen events 1--2 days in advance. We also define the metrics to gauge the performance of the coupling quantitatively. Finally, we summarize and discuss the results in Section~\ref{sec:conclusion}.

%%%%%%%%%%%%%%%%%%%%%%%%%%%%%%%%%%%%%%%%%%%%%%%%%%%%%%%%%%%%%
%
%-------------------------Section---------------------------%
%
%%%%%%%%%%%%%%%%%%%%%%%%%%%%%%%%%%%%%%%%%%%%%%%%%%%%%%%%%%%%%
\section{Models and coupling methodology}
\label{sec:models}

\subsection{EUropean Heliospheric FORecasting Information Asset (EUHFORIA)}
The EUHFORIA architecture comprises a coronal part, which is a 3D semi-empirical model based on the Wang-Sheeley-Arge \cite<WSA,>{arge2004} model, and a heliospheric part. The coronal part uses the photospheric magnetic field from the synoptic magnetogram maps and generates the solar wind plasma conditions using empirical relations at 0.1~au, i.e., the inner boundary of the heliospheric part \cite{Pomoell2018,Asvestari2019}. The heliospheric part, a 3D time-dependent model of the inner heliosphere, numerically solves the ideal MHD equations (including gravity) employing a cell-average finite volume method in Heliocentric Earth EQuatorial (HEEQ) coordinate system. The divergence-free condition is ensured by implementing the Constrained Transport method \cite{Evans1988}. CMEs are inserted into the heliospheric part as time-dependent boundary conditions, and then self-consistently advanced by the MHD equations in the heliosphere. In addition to the cone model \cite<simplified non-magnetized spherical blob of plasma,>{Pomoell2018,Scolini2018}, magnetized CME models ({Linear Force Free (LFF)} spheromak model, \citeA{Verbeke2019}; {Flux Rope in 3D (FRi3D)} model, \citeA{Maharana2022}) are incorporated in EUHFORIA. This makes it possible to predict the magnetic field components in the heliosphere, which mainly determine the severity of the geomagnetic storms.

In this work, we use EUHFORIA (ver 2.0) as installed on the wICE cluster of the Vlaams Supercomputer Centrum ({\url{http://www.vscentrum.be}}). Two nodes of this supercomputer with 72 cores per node (144 parallel processes) were utilized. The computational mesh has a radial resolution of 0.0037~au (corresponding to $0.798\;$R$_\odot$) for 512 cells in the radial direction between 0.1 -- 2~au, and an angular resolution of $2^\circ$ in the latitudinal direction between $\pm80^\circ$ and $2^\circ$ in the longitudinal directions extending between 0--360$^\circ$. This resolution is used at space weather modeling centers like CCMC and VSWMC. EUHFORIA typically provides global 3D MHD output (with hourly cadence, i.e., temporal resolution is one hour) and 1D time series (10-minute cadence) output at any given point in the simulation domain. In the perspective of in situ measurements, the observed solar wind plasma parameters typically have a cadence in the order of a minute, and the magnetic field components data have a cadence in the order of seconds. 
\begin{figure}
\centering
\includegraphics[width=\textwidth,trim={0cm 1.4cm 0cm 0cm},clip=]{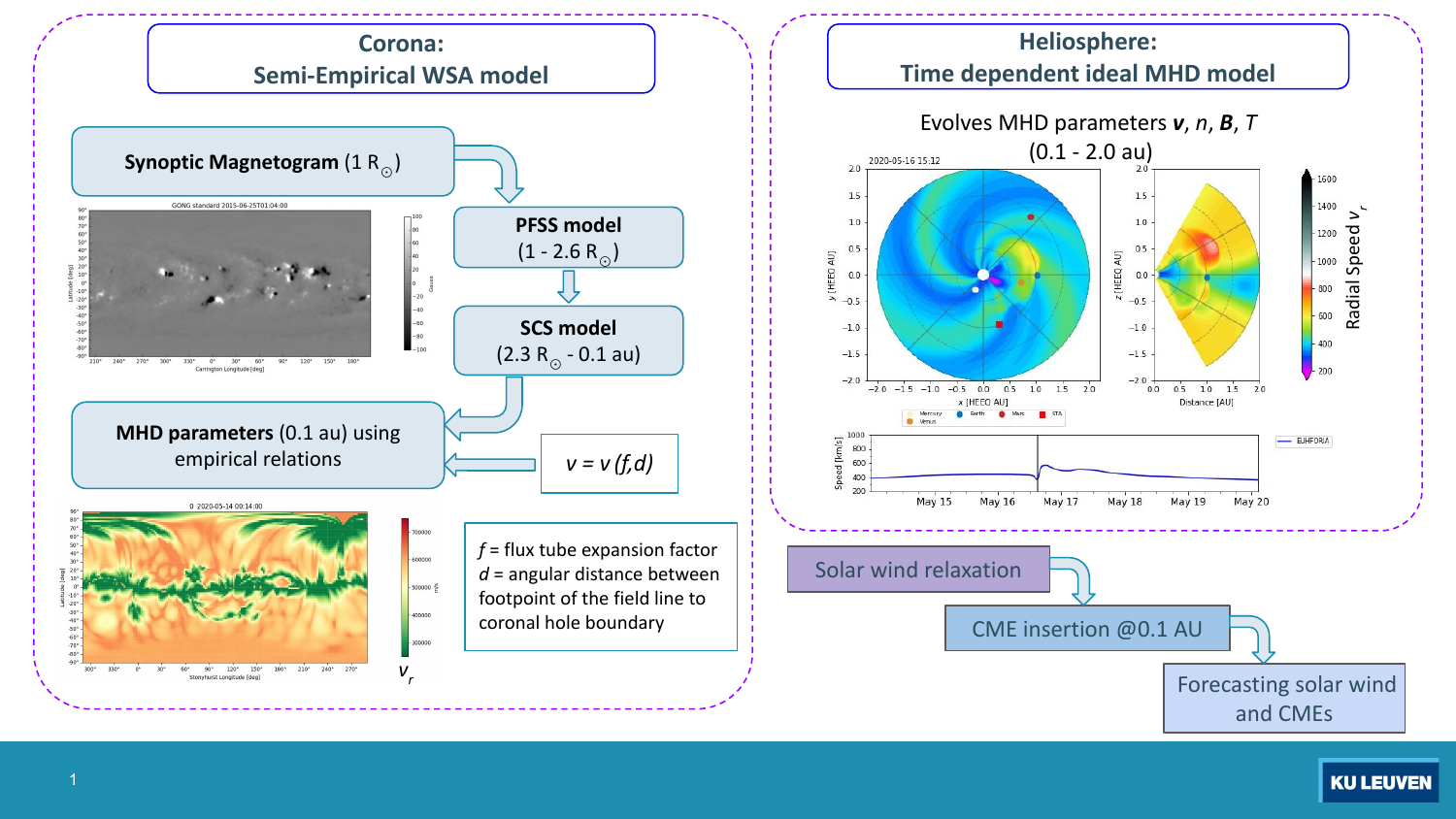}
\caption{Schematic showing the different models within EUHFORIA and the outputs they produce. On the left, the corona is modeled using the semi-empirical Wang-Sheeley-Arge (WSA) model, driven by the synoptic magnetogram maps (e.g., top left figure). WSA model employs the potential field source surface (PFSS) model in the low corona and the Schatten current sheet (SCS) in the upper corona to extrapolate the magnetic field lines up to 0.1~au and employs empirical relations to compute MHD parameters at 0.1~au (e.g., bottom left figure). The output of the coronal model is provided as a boundary condition to the 3D time-dependent ideal MHD model of the heliosphere, shown on the right. Figure on the right is an example showing the radial speed ($v_r$) profile in the EUHFORIA domain depicting a propagating CME on a relaxed solar wind background (equatorial and meridional planes containing Earth).}
\label{fig:euhforia_flowchart}
\end{figure}

\subsection{Open Geospace General Circulation Model (OpenGGCM)} 
The OpenGGCM \cite{Raeder2001a,Raeder2001b} is a global model of Earth's magnetosphere, ionosphere, and thermosphere. It is originally a coupling between the magnetosphere and the global ionosphere-thermosphere model called Coupled Thermosphere Ionosphere Model \cite<CTIM;>{Fuller-Rowell1983,Fuller-Rowell1996,Raeder2001a}. The outer magnetosphere is modeled by solving the semi-conservative form of MHD equations on a 3D stretched Cartesian grid. An explicit second-order predictor–corrector finite difference scheme is used for time stepping. The divergence-free condition is preserved by employing the Constrained Transport method. The MHD grid used in this study is based on \citeA{Cramer2017}. It contains 481 cells spanning $-$35~R$_E$ and 5000~R$_E$ in GSE $x$ coordinate, and 180 cells covering $-$48~R$_E$ to 48~R$_E$ in both GSE $y$ and $z$ coordinates. The minimum cell sizes in the inner magnetosphere are down to 0.167~R$_E$ in $x$ and 0.25~R$_E$ in $y$ and $z$. 
%(NPX, NPY, NPZ) = (13,6,6) \\
%(NX, NY, NZ) = (481,180,180) \\
%Minimum cell size in y and z is 0.25 ($R_E$?)\\
%In $x$-direction, sunward and anti-sunward extensions of the computational domain are 35~R$_E$ and 5000~R$_E$, respectively. In $y$- and $z$- directions, the computational domain extends between $\pm$48~$R_E$.
The electrodynamic coupling of magnetosphere-ionosphere (MI) happens at the inner boundary radius of the ionosphere at 2.1~R$_E$. This coupling is necessary to self-consistently model the closure of field-aligned currents (FACs) generated by the interaction of the solar wind and the magnetosphere, in the ionosphere \cite{Raeder1998}. MI is coupled to the default version of the CTIM, wherein, MI provides the electron precipitation parameters and magnetospheric electric field. CTIM, in turn, provides the conductances and dynamo currents to drive the ionosphere potential in the MI model. 

%[width=0.5\textwidth,trim={7cm 3cm 7cm 4.0cm},clip=]
\begin{figure}[ht]
\centering
\includegraphics[width=0.8\textwidth,trim={0cm 0cm 8cm 0.0cm},clip=]{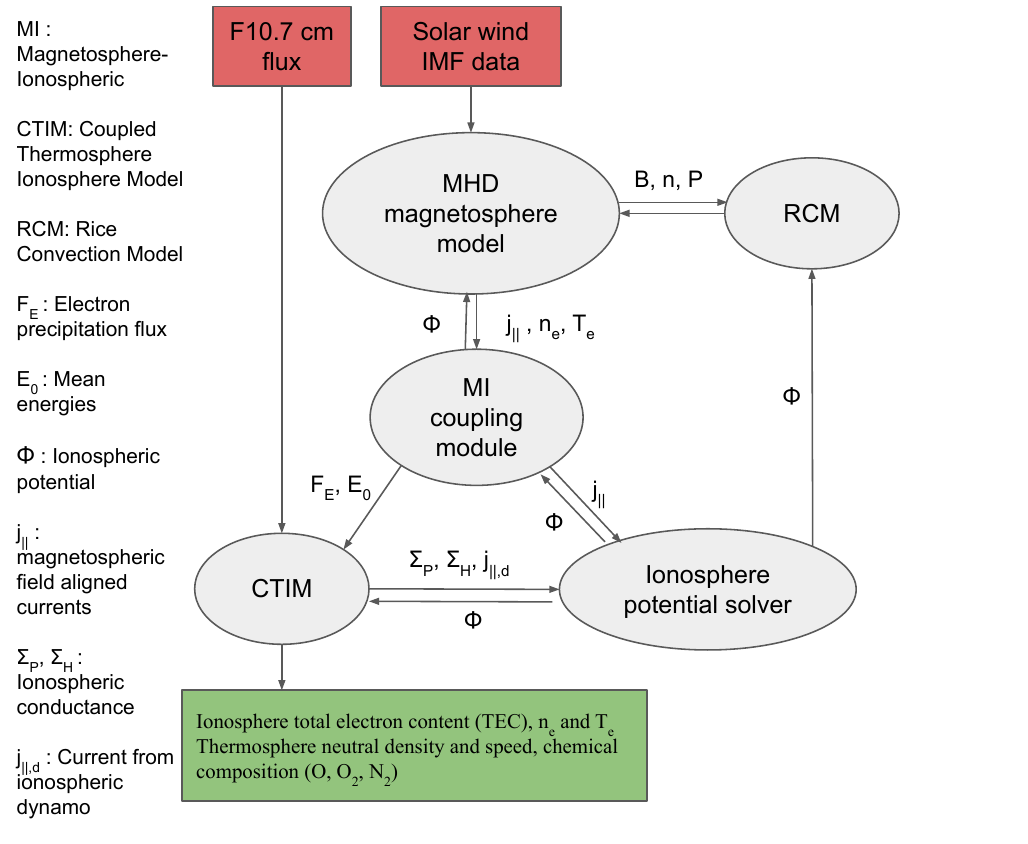}%openggcm_flowchart_rcm.png}
\caption{Flowchart demonstrating the coupling of the MHD magnetosphere model of OpenGGCM with CTIM and RCM, adapted from \citeA{Cramer2017}. B, n, and T are the magnetic field, density, and temperature of the plasma. All other abbreviations and the meanings of the symbols are provided on the left side of the figure.}
\label{fig:openggcm_flowchart}
\end{figure}
As the physics of the inner magnetosphere is not entirely represented by MHD formalism, an additional ring current model is employed for self-consistent feedback. RCM is the ring current model that solves the motion of the plasma flux tubes using the ionosphere potential, magnetic induction, and drift forces. The RCM bounds are at 10~R$_E$ on the dayside, 13~R$_E$ on the nightside, and $\pm$12~R$_E$ on the flanks. %The inner and the outer boundaries of the RCM are located at $\pm$1.02~R$_E$ and $\pm$10~R$_E$, respectively. 
The ring current plasma population is initialized by MHD density and pressure from OpenGGCM. After the execution of RCM, the calculated density and pressure in the RCM domain are fed back into the magnetosphere model to make it converge toward the RCM values. %The MHD pressure and density are slowly adjusted to RCM values. %It is initialized with the external plasma boundary conditions, ionosphere potential, and local magnetic fields. 
The MHD model calculates the precipitation and FACs normally, but the RCM influence has the effect of modifying the precipitation and FAC values indirectly. The calculated precipitation and FACs in the RCM domain are combined with those computed by the MHD magnetosphere model, to in turn drive the CTIM model. %The ionosphere potential is used by the RCM model to derive the precipitation from electron density and FACs from the ring current pressure distribution. 
The physical quantities determined by the RCM model are used at the low latitudes and the MHD quantities at higher latitudes. A simplified schematic representation of the various couplings can be found in Fig.~\ref{fig:openggcm_flowchart}.
%The inner boundary of the RCM is located at 1.02~R$_E$ (1~R$_E$ + ionosphere height = 1 + (120km/RE)) - the height of ionosphere is given by the parameter $RCM_{IONO}$\\
%This simple grid allows an efficient parallelized code as well. The inner boundary of this domain close to Earth is coupled with the ionosphere by the "closure of field-aligned currents (FACs)". 
We use the default setup of OpenGGCM (v5.0.ccmc) with the above specifications, as installed on the Marvin cluster at the University of New Hampshire. %Processors/cores used: NPX*NPY*NPZ + 1 (RCM) + 1 (Ionosphere) + 1 (Diagnostics)\\
%Wall time and CPU time: timestamp(target/\#NAME/*.log) - timestamp(target/\#NAME/*.grid2)\\
We generate 3D MHD output files with an hourly cadence and 2D files at a minute cadence. The ionosphere files are also obtained at a one-minute {cadence} for computing ionospheric geomagnetic indices. The advantage of using this global model is that both the magnetospheric and ionospheric effects of geomagnetic storms can be modeled. The ring current induced in the magnetospheric domain can be used to compute the Dst index. The short timescale changes in the ionosphere at different stations can be used to compute the A-indices, a proxy for substorms. 

\subsection{Coupling EUHFORIA and OpenGGCM}
%In this work, we are interested in a modeling chain to predict geomagnetic indices associated with a CME event. \\
%Earth is a solar-wind dominated MS.

\begin{comment}
\begin{figure*}
\centering
\includegraphics[scale=0.4]{openggcm_coupling.png}
\caption{Flowchart demonstrating the coupling of OpenGGCM with EUHFORIA}
\label{fig:coupling_flowchart}
\end{figure*}
\end{comment}

The global simulation of OpenGGCM is subjected to a time-dependent initialization using solar wind plasma and magnetic field measurements obtained from in situ solar wind satellites. In this study, we couple the results of the heliospheric part of EUHFORIA to OpenGGCM, i.e.\ we provide the output of EUHFORIA at Earth as boundary conditions to drive OpenGGCM. The physical quantities that are supplied are velocity ($\bar{v}$, km/s), magnetic field ($\bar{B}$, nT), proton number density ($n_p$, particles/cc), and thermal pressure ($P$, pPa). The magnetosphere shape and size are driven by the solar wind pressure gradient acting along the normal to the magnetosphere coming from the dynamic pressure ($\rho u^2$, $\rho$ is the mass density and $u$ is the bulk velocity), the plasma pressure ($nk_BT$, $n$ is total number density, $k_B$ is the Boltzmann constant, and $T$ is the plasma temperature), %the kinetic or thermal plasma pressure ($nkT = nkT_{\perp}+nkT_{||}$, $kT_{||} = mv_{||}^2/2$, $T_{||}$ and $T_\perp$ are the temperatures of the particles traversing parallel or perpendicular to the magnetic field, and $kT_\perp = mv_\perp^2$, $n$ = number density, $v_{\perp}$ and $v_{||}$ are the root mean square velocity of a Maxwellian distribution of gyro and parallel velocities respectively), 
and the magnetic field pressure ($B^2/2\mu_0$, $B$ is the IMF magnetic field strength and $\mu_0$ is the magnetic permeability) in the interplanetary {space}. In addition to the normal pressures, tangential stresses transfer momentum and energy across the magnetopause and energize the magnetosphere \cite<Chapter 4; >{Bothmer2007}. All the above physical parameters that are essential to drive the processes in the magnetosphere are provided by the EUHFORIA simulations. 

The standard relaxation time considered for OpenGGCM simulation is around 10--12~hours before the main phase of the storm. Hence, it suffices to start feeding EUHFORIA output to OpenGGCM around 5--6~hours before the CME shock arrival which serves as a relaxation time of around 12~hours for OpenGGCM simulation until the magnetic cloud arrives at Earth. The time cadence of EUHFORIA output is 10 minutes. Given that EUHFORIA time series do not have any fluctuations in the temporal resolution below the conventional time resolution of EUHFORIA (i.e., 10 minutes), we do not increase the input resolution of OpenGGCM.% (the standard input time resolution of OpenGGCM is of the order of 1~minute). %We ensure to initiate the OpenGGCM outer magnetosphere model with low ($\sim$1~nT) but non-zero values of $B_z$. %There is the scope of using the 3D output of OpenGGCM to further couple with GIC models which was pursued under the EUHFORIA 2.0 project.
The output of the coupled model is used specifically for the computation of geomagnetic indices, Dst index, and A-indices. 

%%%%%%%%%%%%%%%%%%%%%%%%%%%%%%%%%%%%%%%%%%%%%%%%%%%%%%%%%%%%%
%
%-------------------------Section---------------------------%
%
%%%%%%%%%%%%%%%%%%%%%%%%%%%%%%%%%%%%%%%%%%%%%%%%%%%%%%%%%%%%%
\section{Geomagnetic indices - measured and modeled}
\label{sec:geomagnetic_indices}
Geomagnetic indices are obtained from ground magnetic field measurements and are proxies for currents in the magnetosphere and the ionosphere, which would be difficult to measure directly. They also have the advantage that they are readily available and that they cover long time periods. In fact, geomagnetic observatories have existed since the 1850s. In this section, the data sources of the geomagnetic indices (Dst and A-indices) and their processing are discussed. We also discuss the methodology to obtain the modeled geomagnetic index data by processing different outputs of the OpenGGCM simulations.

\subsection{Disturbance storm time index (Dst)}
Southward IMF ($B_z$) reconnects with Earth's northward magnetic field and deposits solar wind energy into the magnetosphere. This process can create storm conditions, wherein magnetic disturbances are induced in the magnetosphere and on the ground \cite{Chapman1918}.The Dst index is most sensitive to Earth’s ring current, which flows in the inner magnetosphere. It is obtained from 4 stations near the equator that are nearly evenly spaced in longitude ({\url{https://isgi.unistra.fr/indices_dst.php}}). Dst index data has been available since 1957, but the drawback is its 3-hour cadence. An improved version of this index is the so-called SYM-H index, which is similarly obtained, but with 6 stations and a much better cadence. Data is openly available on the World Data Center (WDC), Kyoto website ({\url{https://wdc.kugi.kyoto-u.ac.jp/wdc/Sec3.html}}). %The SYM-H index is the longitudinally symmetric horizontal (dipole direction) component obtained from the 5-minute averaged magnetic field disturbances at the six geomagnetic observatories (\url{https://wdc.kugi.kyoto-u.ac.jp/aeasy/asy.pdf}). 
In the following, we use SYM-H and Dst interchangeably. It can be shown that the Dst index is also a measure of the plasma energy content of the inner magnetosphere \cite{Dessler1959,Turner2001}. However, the Dst index is also sensitive to other magnetosphere currents, most importantly the currents on the magnetopause. Therefore, the raw Dst index is usually corrected as follows \cite{Burton1975}: 
%Apart from the dominant ring current, the magnetosphere contains many other current systems like the magnetopause currents (Chapman-Ferraro current), field-aligned currents (FAC), and cross-tail currents. SYM-H encompasses the contribution of all types of currents in the domain. The measured SYM-H is simply referred to as Dst hereafter and is corrected 
 %for contamination due to magnetospheric currents as follows:
\begin{linenomath}
\begin{equation}
    \text{Dst}^* =  (\text{Dst} - b \sqrt{P_{dyn}} + c)/a,
    \label{eq:dst_corr}
\end{equation}
\end{linenomath}
where Dst$^*$ is the corrected Dst.
%The raw Dst represents the total equatorial field at Earth's surface produced by all magnetospheric current systems except for the solar wind pressure-induced currents \cite{Burton1975}, 
Here, $P_{dyn}$ is the solar wind dynamic pressure, $a$ is a constant representing the contribution for the ground-induced currents, $b$ [nT/$\sqrt{\mathrm{nPa}}$] is a scaling constant, and $c$ [nT] accounts for the quiet-time contribution. These corrections and other important properties of these indices are discussed in \citeA{Wanliss2006}. Note that an enhanced ring current in the main phase of the storm corresponds to negative Dst {until it reaches a minimum. After that, the recovery phase begins and the Dst values start increasing to restore back to pre-storm conditions (close to zero).}

The model Dst is obtained from integrating the currents in the inner magnetosphere using Biot-Savart’s law to obtain the perturbations at Earth’s surface \cite{Cramer2017}. %This computed Dst is then corrected using Eq.\ref{eq:dst_corr}, and then compared to the observed Dst.
Since the Dst computed in OpenGGCM considers purely the contribution of the ring current (i.e.\ Dst$^*$), we invert Eq.\ref{eq:dst_corr} to obtain Dst from the simulation data to compare it with the measured Dst. 
The empirical constants ($a$, $b$, and $c$) are taken from AK2 model \citeA{Obrien2000a} and their values are $a = 1$, $b=7.26$, $c=11$, since their model resulted in the least root mean square error between the measured and empirically modeled Dst. %The model that resulted in the least RMSE between the observed and modeled Dst was AK2 model \cite{Obrien2000a} with ($a = 1$, $b=7.26$, $c=11$). Hence, we use the coefficients based on the AK2 for incorporating the inverse corrections to the Dst computed by OpenGGCM to match with observations. 

%As the Dst derived from the ground and high-altitude magnetic perturbations may yield contaminations due to other magnetospheric and locally induced currents, computation of Dst by direct measurement of ring current is adopted in the magnetosphere research community \cite{Turner2001}. 
%RCM is employed in OpenGGCM to evaluate the ring current distribution. OpenGGCM traces the closed magnetic field lines in the RCM region wherever non-zero RCM pressure is present \cite{Cramer2017}. 

\subsection{The auroral indices}
The auroral indices or A-indices \cite<AU, AL, and AE, >{Davis1966} are widely used to quantify ionospheric disturbances. They are obtained from 12 stations at auroral latitudes (around 70 degrees magnetic latitude) with more or less even spacing in longitude (See for more details, \url{https://wdc.kugi.kyoto-u.ac.jp/wdc/pdf/AEDst_version_def_v2.pdf}) . The AU index is the upper envelope of the 12 time traces of the north-south perturbations, and, correspondingly, the AL index is the lower envelope. The AE index is the difference, AU-AL. AL is considered to be a measure of the westward electrojet, and thus a good indicator of substorms. By contrast, AU is considered an indicator of eastward currents, which occur mainly on the dayside. Unlike Dst, there is no known correlation between the A-indices and the actual currents. The traditional A-indices are no longer computed on a regular and reliable basis, mostly because data from Russian stations are lacking, which reduces the reliability of the observations used in this work. The SuperMAG magnetometer site, consolidating most worldwide magnetometer measurements, provides an alternative with their SML/SMU/SME indices, which are derived from a different set of stations \cite{Newell2011,dejong2018}.
Unlike Dst, the A-indices also respond to much weaker forcing, i.e., substorms, pseudo-breakups, and Steady Magnetosphere Convection (SMC) events (see, for example, \citeA{Bergin2020}
), which hardly register in Dst data. From the model, we compute the A-indices using Biot-Savart integration over the ionosphere currents. Details of that procedure can be found in \citeA{Raeder2001a,Raeder2001b}.

%%%%%%%%%%%%%%%%%%%%%%%%%%%%%%%%%%%%%%%%%%%%%%%%%%%%%%%%%%%%%
%
%-------------------------Section---------------------------%
%
%%%%%%%%%%%%%%%%%%%%%%%%%%%%%%%%%%%%%%%%%%%%%%%%%%%%%%%%%%%%%
\section{Validation sample}
\label{sec:validation}
In this section, we focus on two events to validate the EUHFORIA-OpenGGCM coupling. Event 1 is a single CME event where the CME was initiated on 12 July 2012 and caused a moderate geomagnetic storm. Event 2 is a more complex event resulting from the interactions between three CMEs that erupted between 4--6 September 2017, resulting in an intense geomagnetic storm at Earth. With Event 1, we aim to validate the capability of OpenGGCM to reproduce a single prolonged main-phase storm, and with Event 2, we extend its capabilities to model multi-step consecutive storms. 
\begin{table}[!t]
\centering
%\begin{tabular}{l{4cm}||c{2.5cm}||c{2.5cm} c{2.5cm} c{2.5cm}}
\begin{tabular}{l || c || c  c  c }
 \hline
 \hline
 \multicolumn{5}{c}{{CME parameters}} \\
 \hline
 Event $\rightarrow$ & Event 1 & \multicolumn{3}{c}{Event 2} \\
 \hline
 Parameters $\downarrow$   & CME1 & CME1 & CME2 & CME3 \\
 \hline
 \hline
 CME model   & spheromak & spheromak & spheromak & spheromak \\
 Insertion date &  2012-07-12  & 2017-09-04  & 2017-09-04   & 2017-09-06 \\
 Insertion time &  19:24  & 22:44  & 23:00   & 14:11 \\
 Speed~[km~s$^{-1}$]  & 763 & 1057 & 697 &  1293\\
 Longitude [$^\circ$]   & $-4$ & $0$ & $25$ &  $21$\\
 Latitude [$^\circ$]    & $-8$ & $-25$ & $0$ &  $-11$  \\
 Radius~[R$_\odot$]       & $16.8$ & $15.2$ & $9.76$ & $11.7$ \\
 Density~[kg~m$^{-3}$]      & $1 \cdot 10^{-18}$ & $1 \cdot 10^{-18}$ & $1 \cdot 10^{-18}$ & $1 \cdot 10^{-18}$ \\
 Temperature~[K]  & $0.8 \cdot 10^6$ & $0.8 \cdot 10^6$ & $0.8 \cdot 10^6$ & $0.8 \cdot 10^6$ \\
 Helicity     & $+1$ & $-1$ & $-1$ & $-1$ \\ 
 Tilt [$^\circ$]        & $135$ & $0$ & $0$ & $40$ \\
 Toroidal magnetic \newline{ flux~[Wb]} & $1 \cdot 10^{14}$ & $4.3 \cdot 10^{13}$ & $4.6 \cdot 10^{13}$ & $7.2 \cdot 10^{13}$ \\
 \hline
 \hline
\end{tabular}
\caption{Parameters of the single CME erupting on July 12, 2012 (Event 1) and of the three CMEs erupting between September 4--6, 2017 (Event 2){ used for initializing CMEs at the EUHFORIA heliospheric inner boundary at 0.1~au.}}
\label{tab:euh_params}
\end{table}

\subsection{Event 1: 12 July 2012}
This is a textbook geoeffective CME event extensively studied in the solar physics and space weather community \cite{Hu2016,Gopalswamy2018,Scolini2019}. The CME erupted on 12 July 2012 from the NOAA AR 11520 located at S17E06. It was associated with an intense GOES X1.4 flare. Based on the source region observations and 3D reconstruction of the CME, it was not expected to impact Earth \cite{Webb2017}. However, a fast (an average projected speed of 885~km~s$^{-1}$), non-interacting halo CME arrived at Earth on 14 July 2012 and resulted in a prolonged negative $B_z$ signature (minimum $B_z=-18$~nT), causing a moderate storm with the minimum $\mathrm{Dst}=-122$~nT on 15 July 2012. The AE index reached a maximum of $\sim$1600~nT with fluctuations close to $1200\;$~nT during the main phase of the storm. The planetary K (Kp) index reached above 5 during the main phase of the storm ({\url{https://www.swpc.noaa.gov/products/planetary-k-index}}). A detailed analysis of the geomagnetic indices threshold (including Dst) by \citeA{Palacios2018} characterizes the severity of this storm towards the end of the moderate category, and close to an intense storm. Due to the prolonged nature of the negative $B_z$, the main phase of the storm started around {10:00}~UT on 15 July 2012 and maintained a $\mathrm{Dst}< -100$~nT until around {09:00}~UT on 16 July 2012. Comparing the in situ measurements from ACE \cite{Chiu1998}, Wind \cite{Ogilvie1997}, and OMNI database ({\url{https://omniweb.gsfc.nasa.gov/}}) during 14--17 July 2012, we found that Wind had the best data, i.e., without any gaps or anomaly. The in situ observations of speed, number density, and $B_z$ are plotted in Fig.~\ref{fig:event1_results}, along with the demarcation of the time of arrival of shock and magnetic cloud. {We choose data sources without gaps and anomaly to facilitate better input for the OpenGGCM simulations}. As per the Wind ICME catalog ({\url{https://wind.nasa.gov/ICME_catalog/}}), the CME shock was recorded on 14 July at $\sim$17:40~UT. The sheath region starting after the shock is characterized by an increased magnetic field, speed, and density. The passage of the magnetic ejecta lasts for almost 2 days between 15 July at $\sim$06:14~UT and 17 July at $\sim$03:22~UT, corresponding to low plasma beta (ratio of the plasma pressure to the magnetic pressure) and proton temperature. %, and a prolonged negative $B_z$ reaching a minimum of around $-18$~nT. 
The $B_z$ component fluctuates between $\pm20\;$nT and stays southward for almost 20~hours, potentially responsible for loading the magnetosphere with energy and reconnecting with Earth's northward magnetic field and erosion. \citeA{Raeder2001a} re-emphasize that $B_z$ is most affecting in the OpenGGCM simulations. 

EUHFORIA simulation was carried out for this event using the magnetized spheromak model \cite{Verbeke2019}. The initial parameters for inserting the CME into the EUHFORIA {heliospheric domain at 0.1~au} are taken from \citeA{Scolini2019}. In this work, we simulate the CME by modifying the magnetic axis orientation, by aligning the magnetic axis of the spheromak with the observed polarity inversion line (PIL) orientation at the source region, to obtain a better fit to the in situ observations as compared to \citeA{Scolini2019}. The CME parameters are provided in Table~\ref{tab:euh_params}. This modification was adopted to improve the modeling of the $B_z$ component. {We have a better match with the minimum negative $B_z$ as compared to the results presented in \citeA{Scolini2019}}. 

\subsection{Event 2: 4--6 September 2017}
This event is associated with the interaction of the three successive CMEs that erupted between 4--6 September 2017 resulting in a two-step geomagnetic storm. It significantly impacted Earth by disrupting telecommunication during hurricane mitigation \cite{Redmon2018}, enhancing radiation dosages \cite{Kataoka2018,Berger2018} and ground-level currents \cite{Cohen2018}, and impacting the ionosphere, GNSS, and radio wave propagation \cite{Yasyukevich2018}.  %This event is a multiple storm case caused due to the complex CME-CME interaction between three successive CMEs that erupted between 4-6 September 2017. 
The geoeffective signatures were observed at Earth between 6--9 September 2017. The first CME (hereafter CME1) was associated with an M1.7 flare that had an average speed of about 600~km~s$^{-1}$. CME1 was overtaken by the faster second CME (hereafter CME2) at $\sim$10~R$_\odot$. CME2 was associated with an M5.5 flare that erupted from the same active region with an average speed of $1420\;$km~s$^{-1}$. CME1 and CME2 merged in the corona, an hour after the eruption of CME2, to drive a single shock. The final full-halo CME (hereafter CME3) erupted on 6 September 2017, associated with an X9.3 flare, and propagated with a projected speed of about $\sim 1570\;$km~s$^{-1}$.

The best in situ observations for this event obtained from the OMNI database (data gaps present in ACE and Wind data) are plotted in Fig.~\ref{fig:event2_results}, along with the demarcation of the time of arrival of shock{, sheath} and magnetic cloud as reported in the Wind ICME catalog. The first shock (hereafter S1) was recorded on September 6 at around 23.00~UT followed by a 0.25~au wide sheath and then a magnetic ejecta associated with the merged CME1 and CME2 (hereafter E1) resulted in a minimum $B_z$ of around $-32\;$nT on September 7, 23.00~UT. E1 was intercepted by the shock of CME3 (hereafter S2) which could have amplified the southward $B_z$ in E1. This enhancement led to the formation of the first and the strongest dip in the Dst index ($-144\;$nT) on September 8 at $\sim$1.00~UT. There are two patches of magnetic ejecta associated with the passage of CME3 (minimum $B_z=-16\;$nT on September 8, 11:00~UT) which led to the development of the second dip in the Dst index (minimum of $-124\;$nT) on September 8 around 13:00~UT. We perform the EUHFORIA simulation of this event using the solar wind and CME parameters (spheromak model) as used by \citeA{Scolini2020}. The CME parameters are listed in Table~\ref{tab:euh_params}. The first dip in B$_z$ is captured by the EUHFORIA simulation, whereas the second dip modeled by the EUHFORIA simulation is underestimated compared to observations.

%%%%%%%%%%%%%%%%%%%%%%%%%%%%%%%%%%%%%%%%%%%%%%%%%%%%%%%%%%%%%
%-------------------------Section---------------------------%
%%%%%%%%%%%%%%%%%%%%%%%%%%%%%%%%%%%%%%%%%%%%%%%%%%%%%%%%%%%%%
\section{Analysis}
\label{sec:analysis}

In this section, we describe the results of the OpenGGCM simulations of the two storms described in Section~\ref{sec:validation}. %The geomagnetic indices are computed using the electric currents and magnetic fields obtained from the global OpenGGCM simulations as described in Section~\ref{sec:geomagnetic_indices}. 
For each of the events, we perform two OpenGGCM simulations - one with the in situ observations recorded at L1 and the other with modeled solar wind output obtained from EUHFORIA simulation. They are named Event`n'-obs and Event`n'-euh, respectively, where `n' is the event number (either 1 or 2 for Event 1 or Event 2, respectively). %We define the Event`n'-obs as a reference simulation, which is then used to compare Event`n'-euh. 
We then compare the geomagnetic activity predicted by OpenGGCM, mainly the Dst and the AE index using the different sets of input. The event-specific Dst and A-indices predictions using OpenGGCM are plotted in Fig.~\ref{fig:event1_results} and Fig.~\ref{fig:event2_results}. For the efficient usage of the computational resources, we choose a narrow window for running OpenGGCM, i.e., an interval corresponding to 6--12 hours before the arrival of the CME until the passage of the magnetic cloud through Earth. 

%\begin{sidewaystable}
%\begin{adjustbox}{angle=90}
%\tabcolsep=2pt

\begin{table*}[!h]
%\Rotatebox{90}{
\begin{tabular}{ l | p{2.cm} | p{1.5cm}  p{1.5cm}  p{1.5cm}  p{1.5cm} }
\hline
\hline
Run ID & Input & Minimum Dst [nT] & Maximum AU [nT] & Minimum AL [nT] & Maximum AE [nT] \\
\hline
\hline
Event1-obs & Wind data & $-151$ \newline$(24\%)$ & 939 \newline$(45\%)$ & $-1667$ $(23\%)$ & 1781 \newline$(0.4\%)$ \\
\hline
Event1-euh & EUHFORIA  \cite{Scolini2019} & $-205$ \newline$(68\%)$ &  945 \newline$(46\%)$ &  $-884$ \newline($-35\%$) & 1800 \newline$(1\%)$\\
\hline
Observations & -- & $-122$ &  648 &  $-1360$ & 1774\\
\hline
\hline
Event2-obs &  OMNI data & $-175$ \newline($22\%$) &  1161 \newline($117\%$) &  $-1136$ \newline($-51\%$) & 1765 \newline($-25\%$)\\
\hline
Event2-euh & EUHFORIA  \cite{Scolini2020} & $-156$ \newline$(8\%)$ &  670 \newline$(25\%)$ &  $-1300$ \newline($-44\%$) & 1825 \newline($-22\%$)\\
\hline
Observations & -- & $-144$ &  535 &  $-2339$ & 2351\\
\hline
\hline
\end{tabular}
\caption{Specifications of the EUHFORIA runs of Event 1 and Event 2 used for validating the EUHFORIA and OpenGGCM coupling. In column 1, Run ID defines an identifier of each simulation, and the input column specifies the source that drives OpenGGCM. The extrema of Dst, AU, AL, and AE indices extracted from the simulations are provided in columns 3--6. The corresponding relative errors (in percentages) of the values obtained from OpenGGCM simulations compared observations are accompanied in the brackets.}
\label{tab:dst_ae_extreme}
%}
\end{table*}
%\end{adjustbox}
%\end{sidewaystable}
For assessing the prediction of the Dst index, we choose to evaluate a quite extended period before and after the Dst dip, instead of just evaluating its single minimum value. To do that, we adopt the dynamic time warping \cite<DTW, >[and references therein]{Gorecki2013,Keogh2001} technique, a method that can help us assess the overall performance of the predicted Dst time series compared to the observed data. DTW measures the similarity between two sequences that have similar patterns but differ in time. The sequences are warped in a non-linear manner to match each other. The optimal alignment of time-dependent sequences is achieved by finding an optimal path through the DTW cost matrix such that the cumulative cost is minimal compared to the other possible paths  \cite{Keogh2001}. This method has been adopted in the space weather community for comparing solar wind characteristics \cite{Samara2022} and Dst index as well  \cite{Laperre2020}. % ``is obtained by finding a path through the DTW cost matrix that minimizes the total cumulative cost among all other possible paths"  %The algorithm copes with time deformations.
For the application of DTW, we ensured that the initial and final points of the two sequences were aligned and verified the forward mapping of temporally advancing points. We further checked that there were no data gaps in our time series as required for the correct application of the DTW algorithm. We smoothed the higher cadence observed data to make them consistent with the low cadence modeled data. Windowing is applied to restrict the matching of the points within a certain time interval in order to minimize singularities, i.e., a scenario when a single data point in one sequence is mapped to a larger subsection of points in the other sequence. {The window length can vary depending on the temporal spread of the features being aligned in a particular event}. We first apply DTW between the measured Dst and the predicted Dst time series modeled by OpenGGCM. To do that, 
we calculate the DTW cost matrix based on the following equation: 

\begin{equation}
\centering
    D(i,j) = \delta(s_{i},q_{i})+\hbox{min}\{D(i-1, j-1), D(i-1, j), D(i, j-1)\}
\label{eq:DTW}
\end{equation}

\noindent where $D(i,j)$ is the cumulative DTW cost or distance, and $\delta(s_{i},q_{i}) = |s_{i} - q_{i}|$ corresponds to the Euclidean distance between the point $s_{i}$ from one time series and the point $q_{i}$ from the other time series. The first element of the array $D(0,0)$ is equal to $\delta(s_{0},q_{0})$. The last element of this cost matrix is called the DTW score, and can be presumed as a quantification of alignment between the two time series. 
%The DTW score is the last element of the cost matrix and reflects the distance measure between modelled and observed series. 

To properly evaluate the performance of the predicted Dst time series, we compute the sequence similarity factor (SSF), a skill score to evaluate the performance of the predicted Dst sequences based on the ideal (observations) and a reference scenarios. In our case, the reference prediction scenario reflects a worst-case scenario in which no Dst prediction was made by the model, namely, Dst is null (i.e., no geomagnetic disturbances at all). The SSF is the ratio between the DTW score of the observed and modeled Dst time series, and the DTW score between the observed and reference (null) scenario time series. Namely, it is defined as:

%The two Dst predictions from OpenGGCM (Event`n'-obs and Event`n'-euh), are also compared against a reference prediction scenario which, in our case, reflects a worst-case scenario in which no Dst prediction was made by the model. Namely, for benchmarking the analysis in the context of Dst, we define the reference case when Dst is null, i.e., no geomagnetic disturbances at all. Then, we compute the sequence similarity factor (SSF), a skill score to evaluate the performance of the predicted Dst sequences based on the ideal (observations) and the reference (null Dst) scenarios. 

\begin{linenomath}
\begin{equation}
    \text{SSF} = \frac{DTW_{score}(O, M)}{DTW_{score}(O, N)}, \ \text{SSF} \in [0, \infty)
\end{equation}
\end{linenomath}
where $O$, $M$, and $N$ represent the observed, modeled, and null Dst scenario (reference case) respectively. %For Event1, SSF is lower for the OpenGGCM simulation driven by observations - it performs better
%This condition is a non-ideal scenario because the desirable features in the time series that are being aligned and assessed correspond to negative Dst. 
Comparing the SSF of Event`n'-obs and Event`n'-euh, the lower the SSF, the better the prediction is, compared to observations. We also estimate the time differences ($\Delta$t) and amplitude differences ($\Delta$Dst) based on the points that were aligned by DTW between the observed and modeled Dst time series {($\Delta$ refers to the difference between Observed and Modeled values)}. Their distributions are presented in Figs.~\ref{fig:Event1-obs_dst_dtw}, \ref{fig:Event1-euh_dst_dtw}, \ref{fig:Event2-obs_dst_dtw} and \ref{fig:Event2-euh_dst_dtw}. 

\begin{table}
\begin{center}
\begin{tabular}{l    c    c    c}
\hline
\hline
Event & DTW$_{score}$($O$, $M$) & DTW$_{score}$($O$, $N$) & SSF\\
\hline
\hline
Event1-obs & 15076 & 37391 & 0.40\\
\hline
Event1-euh & 17040 & 37391 & 0.45\\
\hline
\hline
Event2-obs & 21117 & 30823 & 0.68\\
\hline
Event2-euh & 13945 & 30823 & 0.45\\
\hline
\hline
\end{tabular}
\end{center}
\caption{DTW scores for each OpenGGCM simulation and their sequence similarity factor (SSF)}
\label{tab:dtw}
\end{table} 

\subsection{Event 1}
\begin{figure}
\centering
\includegraphics[width=\textwidth]{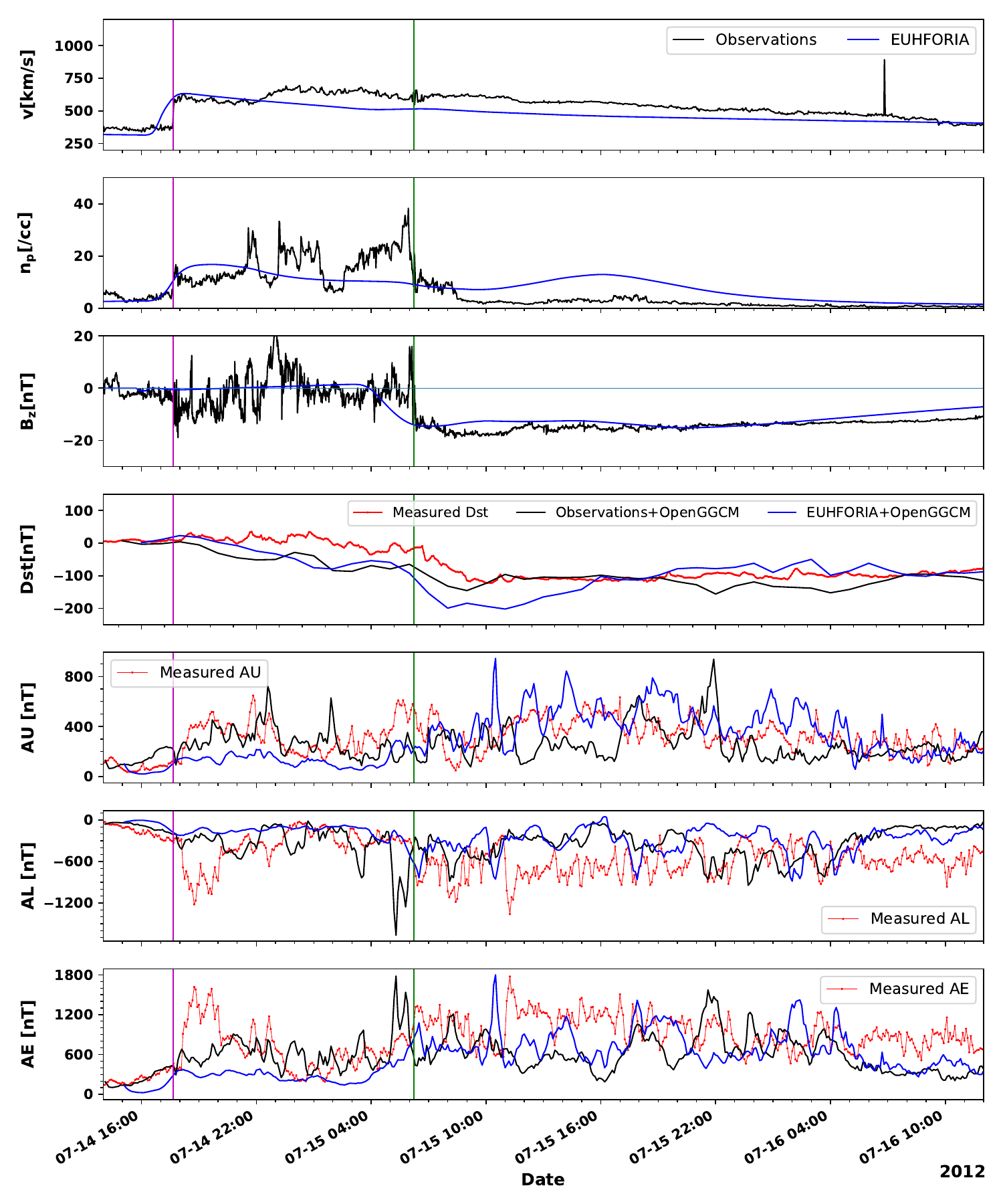}%espm_20170904og.png} 
\caption{Characteristics and predicted geomagnetic indices of Event 1. Panels 1-3 show the plasma parameters -- speed ($v$), proton number density ($n_p$), and the magnetic field parameter -- z-component of magnetic field ($B_z$) as obtained from the Wind spacecraft in situ observations (in black) and the EUHFORIA simulation of the event based on {modified} \citeA{Scolini2019} (in blue), respectively. The horizontal blue line in Panel 3 corresponds to $B_z=0$. Panels 4--7 show the geomagnetic indices -- Dst index, AU index, AL index, and AE index as measured in Earth's magnetosphere and ionosphere (in red), and as obtained from OpenGGCM simulations using input from the Wind (in black) and EUHFORIA simulation (in blue). The magenta and green vertical solid lines depict the arrival of the CME shock and the beginning of the magnetic cloud passage at Earth, respectively.}
\label{fig:event1_results}
\end{figure}

\begin{figure}
\centering
%\subfloat[]{\includegraphics[width=0.9\textwidth]{lia_win_dst_20120712_wi_lowcad_init14h_re.pdf}} 
%\vspace{-1\baselineskip}
%\subfloat[]{\includegraphics[width=0.45\textwidth]{lia_win_hist_tdiff_dst20120712_wi_lowcad_init14h_re.pdf}}
%\subfloat[]{\includegraphics[width=0.45\textwidth]{lia_win_hist_ampdiff_dst20120712_wi_lowcad_init14h_re.pdf}}
{\includegraphics[width=\textwidth]{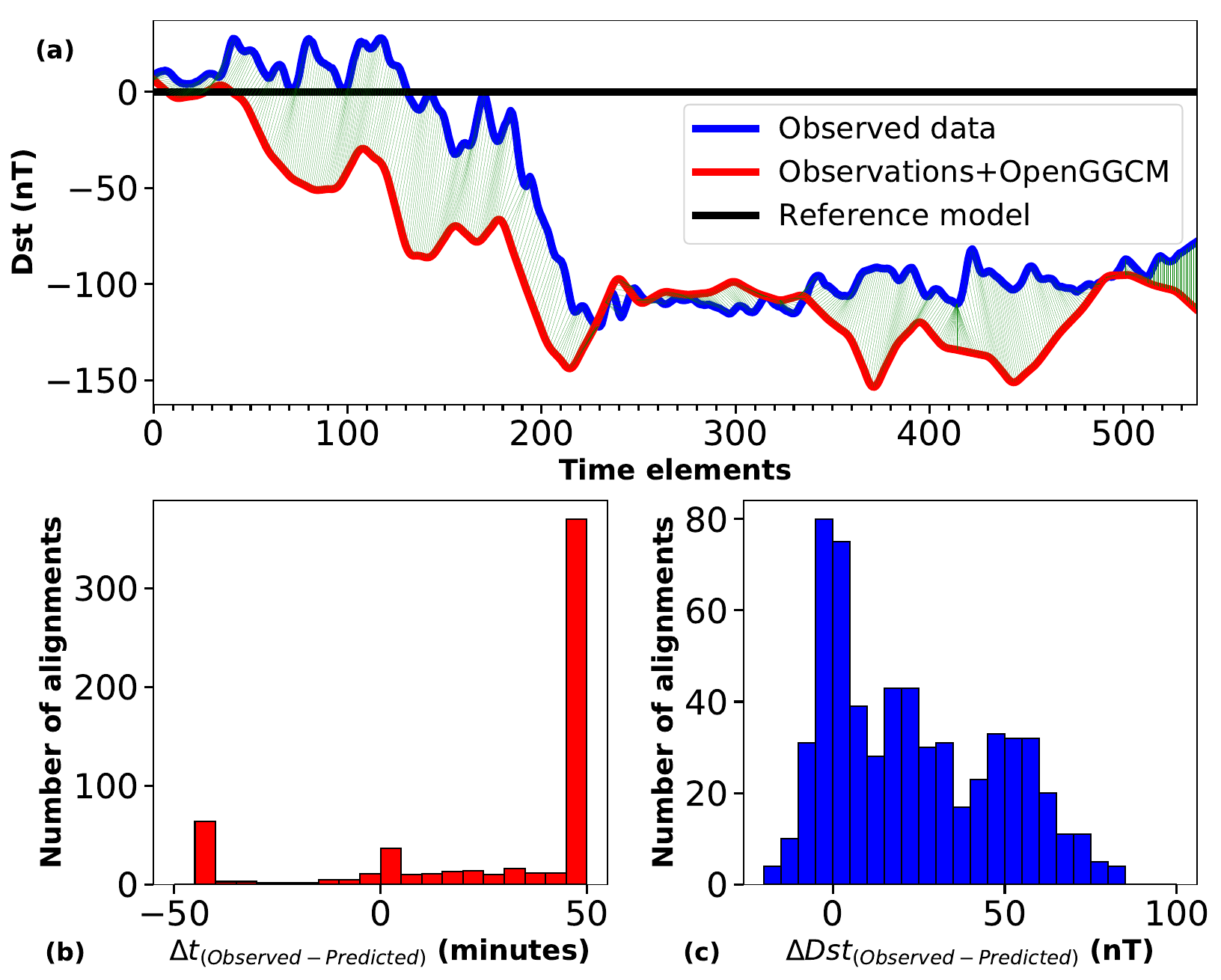}}
\vspace{-1\baselineskip}
\caption{DTW analysis of the Dst index. (a) DTW alignment {between the observed (blue) and predicted/modeled (red) time series}, (b) Histogram of time differences between the aligned points, and (c) Histogram of Dst differences between the aligned points for Event1-obs. Each time element in (a) corresponds to 5-minutes.}
\label{fig:Event1-obs_dst_dtw}
\end{figure}

\begin{figure}
\centering
%\subfloat[]{\includegraphics[width=\textwidth]{lia_win_dst_20120712_lowcad_init15h_-1nT_1.pdf}}
%\vspace{-1\baselineskip}
%\subfloat[]{\includegraphics[width=0.5\textwidth]{lia_win_hist_tdiff_dst20120712_lowcad_init15h_-1nT_1.pdf}}
%\subfloat[]{\includegraphics[width=0.5\textwidth]{lia_win_hist_ampdiff_dst20120712_lowcad_init15h_-1nT_1.pdf}}
{\includegraphics[width=\textwidth]{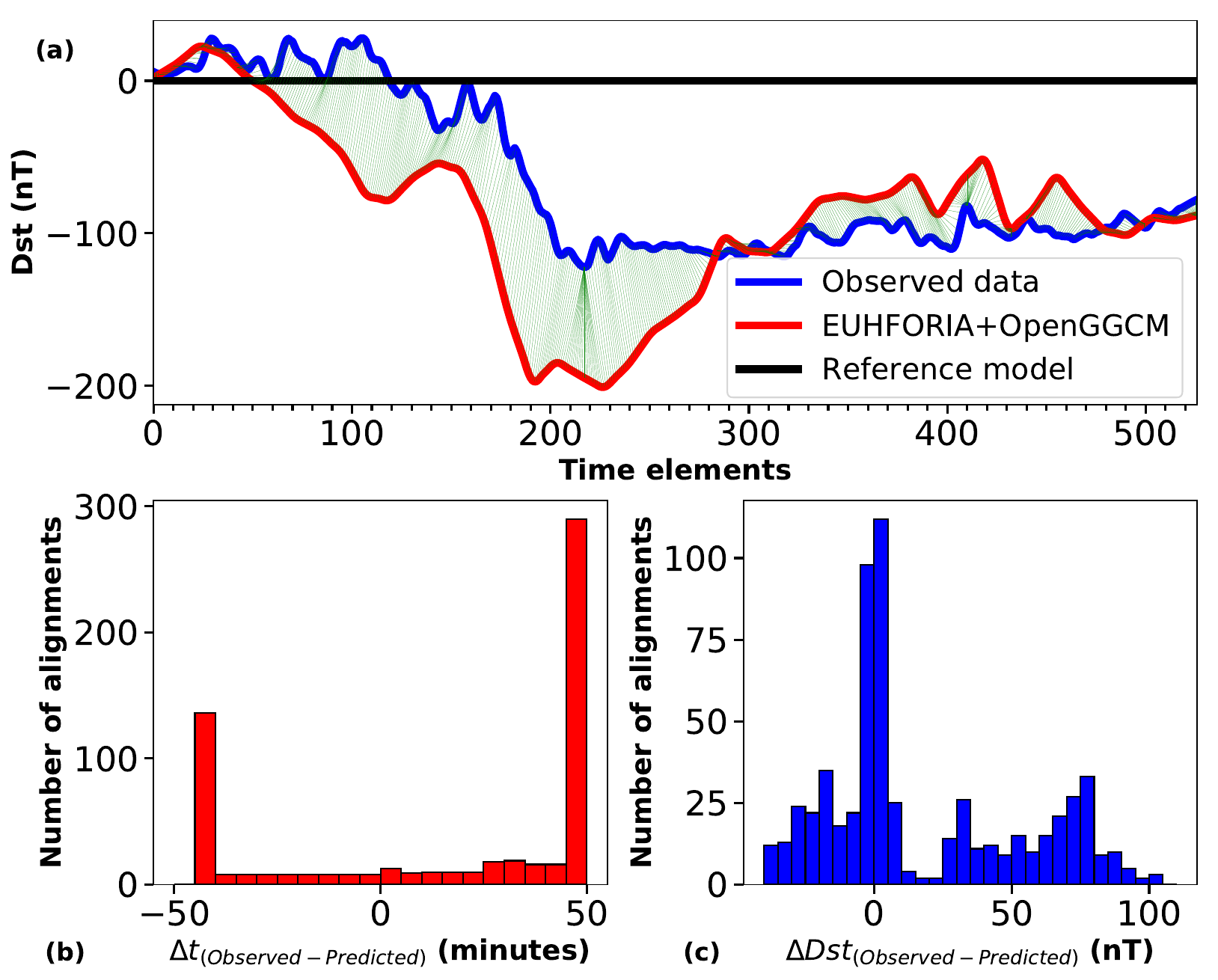}}
\vspace{-1\baselineskip}
\caption{DTW analysis of the Dst index. (a) DTW alignment {between the observed (blue) and predicted/modeled (red) time series}, (b) Histogram of time differences between the aligned points, and (c) Histogram of Dst differences between the aligned points for Event1-euh respectively. Each time element in (a) corresponds to 5-minutes.}
\label{fig:Event1-euh_dst_dtw}
\end{figure}

The inputs, as described in Section~\ref{sec:geomagnetic_indices}, and the results (Dst and A-indices predictions) of the OpenGGCM simulations are presented in Fig.~\ref{fig:event1_results}. %We first discuss the modeling of the Dst index with OpenGGCM. 
The {sudden commencement (Dst starts to become negative), main (rapid decrease in Dst) and recovery (Dst increases to restore back to normal)} phases of the storm are qualitatively modeled in both Event1-obs and Event1-euh. %Rapid fluctuations in the predicted Dst profiles are observed almost midway through the recovery phase (starting around 15 July at 22:00) as a result of certain instabilities towards the end of the OpenGGCM simulations. 
Two important aspects of comparing the Dst profile are the magnitude of the minimum Dst and the time of the beginning of the main phase. In Event1-euh the minimum Dst value is overestimated as compared to Event1-obs, and the main phase is initiated $\sim$2~hours earlier as compared to the observations. This corresponds to about 2~hours of early modeling of minimum $B_z$ in the EUHFORIA simulation compared to the observed profiles (panel 3 in Fig.~\ref{fig:event1_results}). For the application of DTW, a window of 50 minutes was considered based on the time differences between features of the modeled and observed Dst for Event 1. The alignments between the predicted Dst of Event1-obs and Event1-euh with observations is shown in Fig.~\ref{fig:Event1-obs_dst_dtw}(a) and \ref{fig:Event1-euh_dst_dtw}(a), respectively. For both Event1-obs and Event1-euh, most of the alignments showed a $\Delta$t of $+50$~minutes between observed and modeled time series (see Fig.~\ref{fig:Event1-obs_dst_dtw}(b) and \ref{fig:Event1-euh_dst_dtw}(b)). This means that, in most of the cases, the prediction of the Dst profile was earlier compared to observations. Figures~\ref{fig:Event1-obs_dst_dtw}(c) and \ref{fig:Event1-euh_dst_dtw}(c) show the histograms of Dst amplitude difference between observed and modeled Dst time series. In both plots we notice that the $\Delta$Dst is mostly positive, ranging between [-20, 85] nT and [-40, 105] nT, for Event1-obs and Event1-euh, respectively. This indicates that the predicted Dst, for most of the alignments, was more negative compared to observations. The SSF for Event1-obs is slightly lower than Event1-euh (cf. Table~\ref{tab:dtw} for more details). This implies that the OpenGGCM predictions made with the measured solar wind properties can reproduce the observed Dst slightly better, although not significantly better, as compared to the predictions using EUHFORIA simulated data. 

The AE index is a proxy of the response of the ionosphere to the substorms which are quite stochastic and have high temporal variability. The predicted AU, AL, and AE indices match the order of magnitude and show rapid variations during the period of the storm. Although there is a reasonable agreement in the order of magnitude of the AE index, it is not straightforward to align the peaks of the predicted AE indices to their measured counterparts to assess the performance of OpenGGCM. The extreme Dst, AU, AL, and AE indices measured and predicted by Event1-obs and Event1-euh, are provided in Table~\ref{tab:dst_ae_extreme}.
% A maximum $\Delta$t of $-$50~minutes is recorded in both cases (Fig.~\ref{fig:Event1_dst_dtw}(c) and \ref{fig:Event1_dst_dtw}(f)) which corresponds to the {early} prediction of the Dst profile compared to the actual measurements. Figures~\ref{fig:Event1_dst_dtw}(b) and \ref{fig:Event1_dst_dtw}(e) show the highest occurrence of $\Delta$Dst at $-$5~nT and 5~nT, respectively, implying a decent agreement of the sequences on an average. However, the next highest values $\Delta$Dst are around 20~nT and 50~nT for Event1-obs which correspond to the offsets at the significant dips in the modeled time series. In case of Event1-euh, the secondary peaks in $\Delta$Dst appear around $-$20~nT and 70~nT.  %The alignment of Event1-euh with the measurements is presented in Fig.~\ref{fig:Event1_dst_dtw}(d). The maximum $\Delta$t is 10~hours (\ref{fig:Event1_dst_dtw}(e)) similar to Event1-obs corresponding to the beginning phase of the time series. Although $\Delta$Dst peaks at 0~nT (\ref{fig:Event1_dst_dtw}(f))

\begin{comment}
%https://docs.google.com/document/d/14DNWOJlV_cMGvglRF3utstf5PQ0TNk1orwNQl7JEeFw/edit
\begin{figure}%[!ht]
\centering
{\includegraphics[width=\textwidth,trim={2.5cm 2.3cm 2.5cm 0cm},clip=]{event1_dtw_final_600dpi.pdf}}
\caption{DTW analysis of the Dst index. (a,d) DTW alignment {between the observed (blue) and predicted/modeled (red) time series}, (b,e) Histogram of Dst differences between the aligned points, and (c,f) Histogram of time differences between the aligned points for Event1-obs and Event1-euh, respectively. Each time element in (a,d) corresponds to 5-minutes.}
\label{fig:Event1_dst_dtw}
\end{figure}
\end{comment}

\subsection{Event 2}
\begin{figure}%[!ht]
\centering
\includegraphics[width=\textwidth]{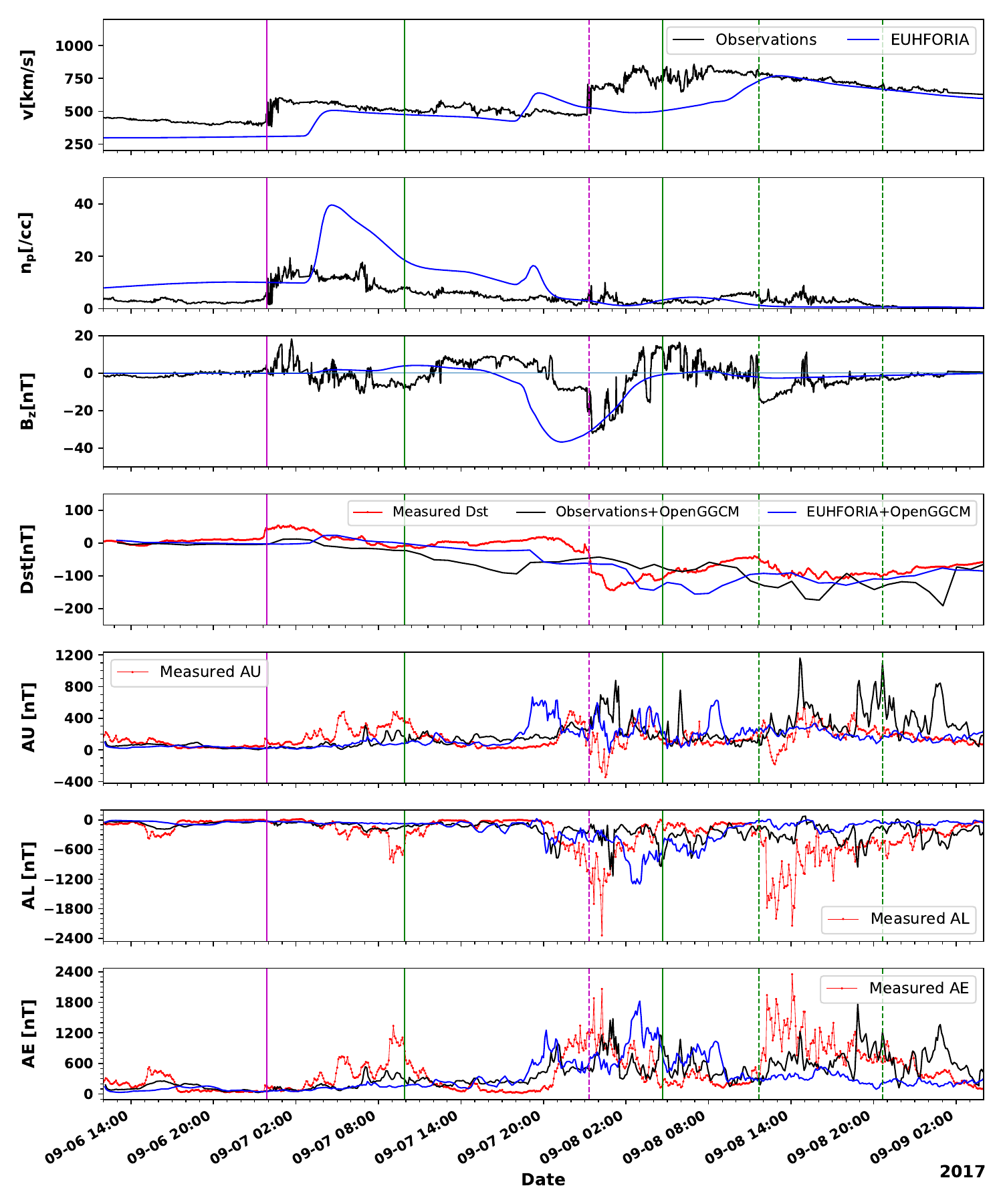}%20170904.png}%espm_20170904og.png}
\caption{Characteristics and predicted geomagnetic indices of Event 2. Panels 1-3 show the plasma parameters -- $v$, $n_p$, and $B_z$ as obtained from the in situ observations {in the OMNI database }(in black) and the EUHFORIA simulation of the event based on  \citeA{Scolini2020} (in blue), respectively. Panels 4-7 show the geomagnetic indices -- Dst index, AU index, AL index, and AE index as measured in Earth's magnetosphere and ionosphere (in red), and as obtained from OpenGGCM simulations using input from the OMNI database (in black) and EUHFORIA simulation (in blue). The magenta solid and dashed lines depict the arrival of two shocks (S1 and S2) associated with this event. The two green solid lines depict the boundary of the passage of the magnetic ejecta E1 at Earth and the dashed lines correspond to the boundary of E2 at Earth. {These interplanetary CME features are detected $40\;$minutes later at Earth (as per OMNI database) relative to L1 (as per WIND ICME catalog).}} %{There is a mismatch between the demarcations by vertical lines and the features in the plot. The vertical lines are based on the identification of the boundaries of the features at L1 as per Wind catalog. Whereas, the plotted observations are from the OMNI database where the data is shifted by $\sim$1~hour to provide the data exactly at Earth's location.}}
\label{fig:event2_results}
\end{figure}

\begin{figure*}%[!ht]
\centering
%\subfloat[]{\includegraphics[width=\textwidth]{lia_win_dst_20170904_omni_lowcad3.pdf}} 
%\vspace{-1\baselineskip}
%\subfloat[]{\includegraphics[width=0.5\textwidth]{lia_win_hist_tdiff_dst20120712_lowcad_init15h_-1nT_1.pdf}}
\subfloat[]{\includegraphics[width=1\textwidth]{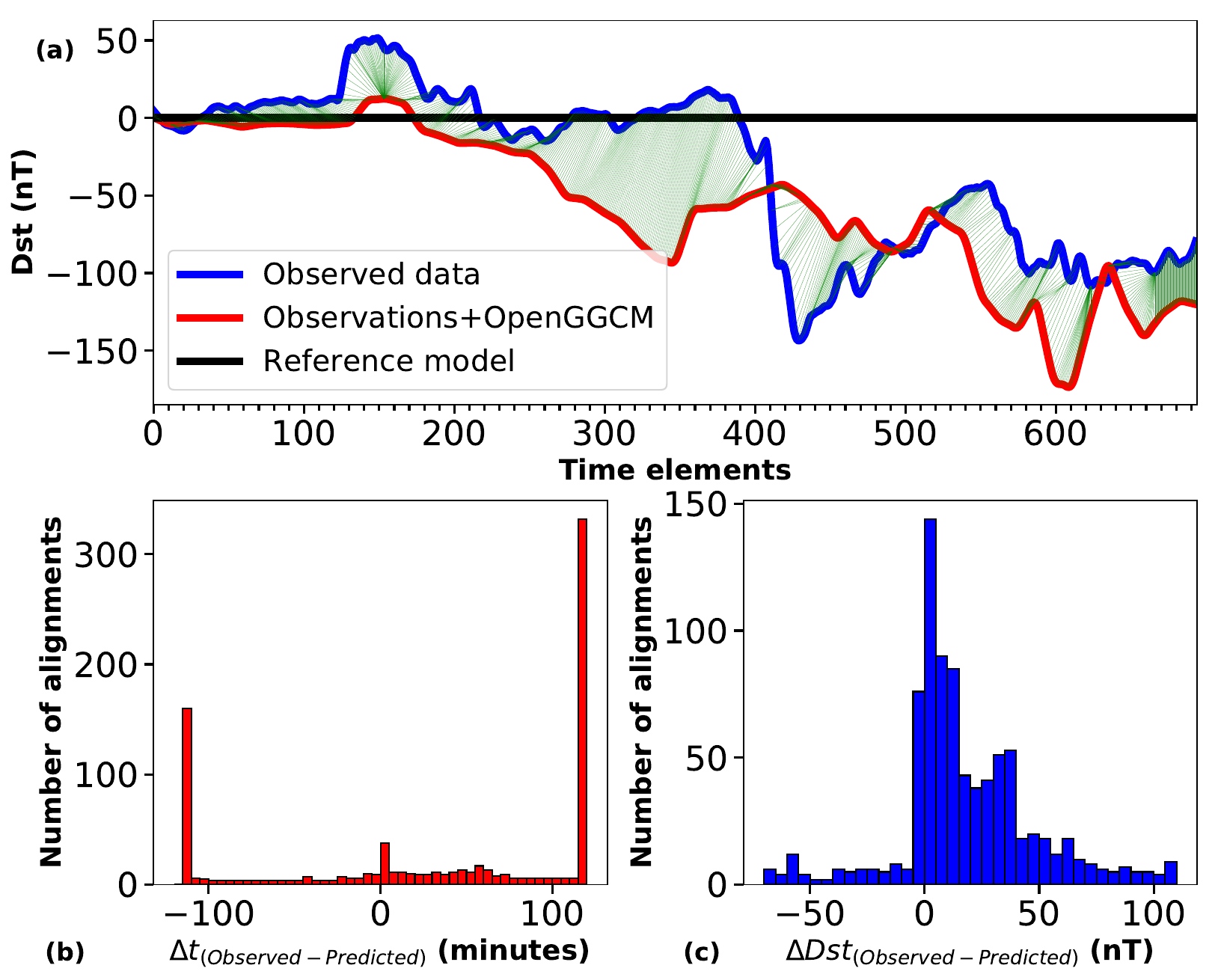}}
\vspace{-1\baselineskip}
\caption{DTW analysis of the Dst index. (a) DTW alignment {between the observed (blue) and predicted/modeled (red) time series}, (b) Histogram of time differences between the aligned points, and (c) Histogram of Dst differences between the aligned points for Event2-obs. Each time element in (a) corresponds to 5-minutes.}
\label{fig:Event2-obs_dst_dtw}
\end{figure*}

\begin{figure*}%[!ht]
\centering
%\subfloat[]{\includegraphics[width=\textwidth]{lia_win_dst_20170904_lowcad_11.pdf}} 
%\vspace{-1\baselineskip}
%\subfloat[]{\includegraphics[width=0.5\textwidth]{lia_win_hist_tdiff_dst20170904_lowcad_11.pdf}}
\subfloat[]{\includegraphics[width=1\textwidth]{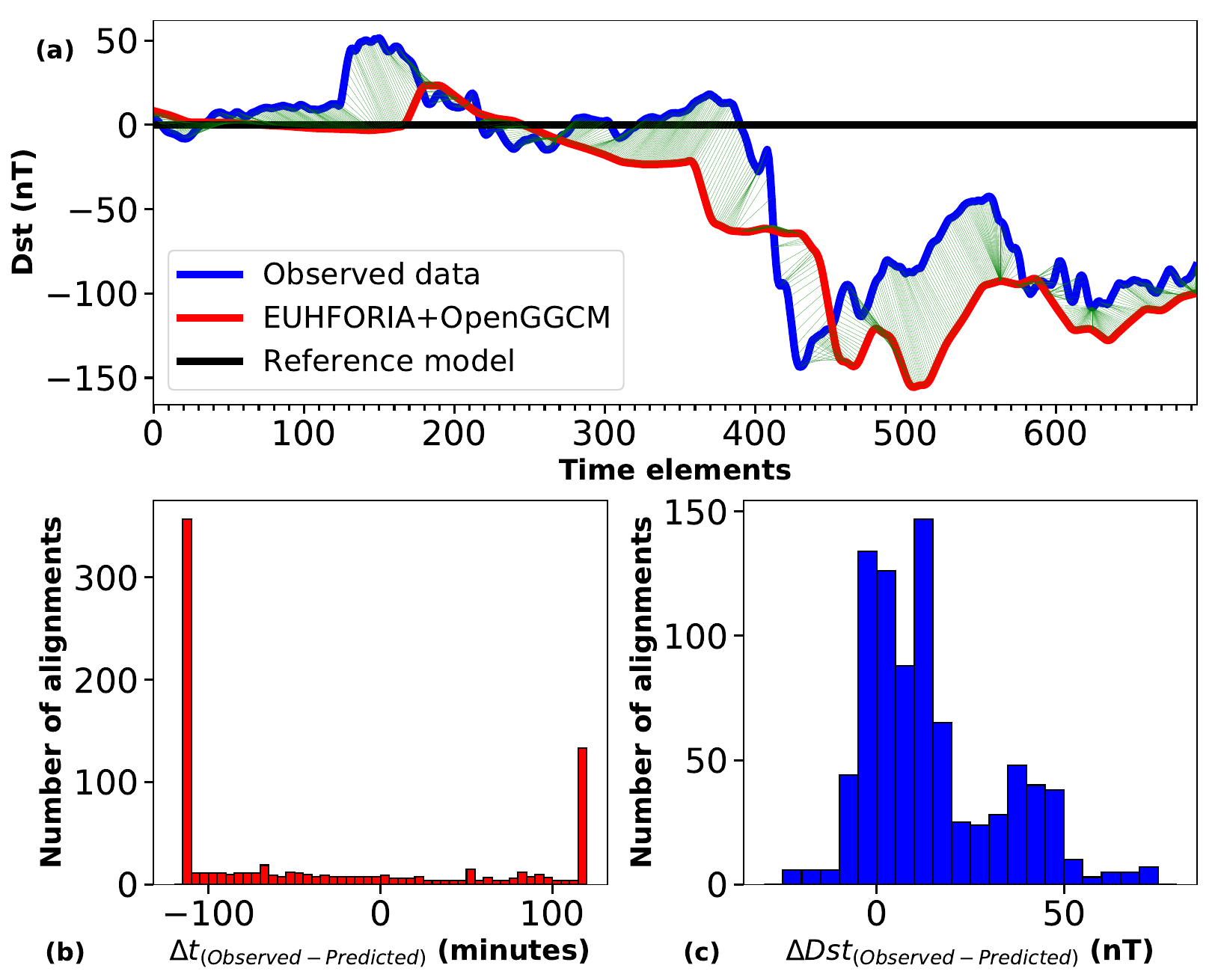}}
\vspace{-1\baselineskip}
\caption{DTW analysis of the Dst index. (a) DTW alignment {between the observed (blue) and predicted/modeled (red) time series}, (b) Histogram of time differences between the aligned points, and (c) Histogram of Dst differences between the aligned points for Event2-euh respectively. Each time element in (a) corresponds to 5-minutes.}
\label{fig:Event2-euh_dst_dtw}
\end{figure*}

The comparison of the predicted Dst and AE indices with the measurements is presented in Fig.~\ref{fig:event2_results}. For this case, the Dst prediction of Event2-euh is qualitatively closer to the measurements as compared to Event2-obs. The two-step feature of the storm is well reproduced by Event2-euh. Event2-obs underestimates the first dip of the Dst. {It is a possibility that the first measured Dst dip (8 September at 01:00) is predicted too early in Event2-obs on 7 September around 18:00. Otherwise, the first dip in Event2-obs Dst could be the overestimated Dst due to the slight negative $B_z$ occurring on 7 September around $\sim$07:00 -- 10:00.} {The DTW window of 120 minutes was applied based on the time differences between features of the modeled and measured Dst for Event2.} The DTW alignments of Event2-obs and Event2-euh with the measured Dst values are shown in Fig.~\ref{fig:Event2-obs_dst_dtw}(a) and \ref{fig:Event2-euh_dst_dtw}(a), respectively. Figures~\ref{fig:Event2-obs_dst_dtw}(b) and \ref{fig:Event2-euh_dst_dtw}(b) present the histograms of time differences between the alignments. In case of Event2-obs, most of the alignments indicate a time difference of 115 to 120~minutes, meaning that the Dst patterns were mostly predicted earlier than observed. The opposite is recorded for Event2-euh. Namely, Fig.~\ref{fig:Event2-euh_dst_dtw}(b) shows that for the majority of alignments, $\Delta$t is negative and equal to $-$110 to $-$115 minutes (most Dst patterns were predicted later than observed). Figures~\ref{fig:Event2-obs_dst_dtw}(c) and \ref{fig:Event2-euh_dst_dtw}(c) present the histograms of the Dst amplitude difference between measured and predicted Dst, for Event2-obs and Event2-euh, respectively. Same as in Event 1, we notice that in both plots $\Delta$Dst is positive for most of the alignments, meaning that the predicted Dst index was mostly overestimated (i.e., more negative) compared to observations. The SSF for Event2-euh is clearly lower than Event2-obs indicating better Dst prediction by OpenGGCM using EUHFORIA input (cf. Table~\ref{tab:dtw} for more details). 

The AE index has two distinct regions of enhancements on 8 September 2017, first around 00:00~UT and the other around {16:00}~UT. Qualitatively, the first peak is modeled by both Event2-obs and Event2-euh, whereas the second peak is better modeled by Event2-obs. Due to high temporal variability, we could not perform a sequence alignment with DTW on the AE time data for this event for the same reasons as mentioned for Event 1.

%The highest occurring $\Delta$t is 120~minutes for Event2-obs (Fig.~\ref{fig:Event2_dst_dtw}(c)) suggesting an early modeling of the features as compared to the observations. For Event2-euh, the highest occuring $\Delta$t is at $-$120~minutes (\ref{fig:Event2_dst_dtw}(f)), suggesting a late occurrence of the majority of features modeled in the simulations. %, especially at the beginning of the time series. 

\begin{comment}
%https://docs.google.com/document/d/1fK8l8lsaY2fhWP1K1IwmGIKuCCgmGIHWqsxYv0644tM/edit
\begin{figure}%[!ht]
\centering
{\includegraphics[width=\textwidth,trim={2.5cm 2.7cm 2.5cm 0cm},clip=]{event2_dtw_final_600dpi.pdf}}
\caption{DTW analysis of the Dst index {between the measured (blue) and predicted/modeled (red) time series}. (a,d) DTW alignment, (b,e) Histogram of Dst differences between the aligned points, and (c,f) Histogram of time differences between the aligned points for Event2-obs and Event2-euh, respectively. Each time element in (a,d) corresponds to 5-minutes.}
\label{fig:Event2_dst_dtw}
\end{figure}    
\end{comment}

%The highest occurring $\Delta$t for Event2-euh is -110~minutes (around -2~hours) corresponding to the delay of both dips in Dst corresponding to the intense and the moderate storms during the period of 8-9 July 2017. 

%The most frequently occurring $\Delta$Dst are 0-5~nT and 10-15~nT respectively (cf. Figures~\ref{fig:Event2_dst_dtw}(b) and \ref{fig:Event2_dst_dtw}(e)), implying a decent agreement on an average. The second highest occurring $\Delta$Dst is around 30-50~nT for both Event2-obs and Event2-euh, corresponding to the offsets at the main phase of the storm. 

\subsection{Sources of error}
%\subsection{Limitations of EUHFORIA}
The first source of error enters through the process of constraining the CME and solar wind parameters for the EUHFORIA simulations. 3D reconstruction of CMEs is subject to human bias  \cite{VERBEKE2022}. Methods of extracting magnetic field parameters for eruptions are dependent on its source region signatures which can be missing in the cases of ``problem storms" \cite{dodson1964}. The errors in CME arrival time and magnetic field predictions are also associated with the ambient solar wind through which the CME propagates. The solar wind maps are limited by the availability and the quality of the synoptic magnetogram, and the semi-empirical WSA model which is one of the simpler coronal models. All these factors determine the accuracy of the predicted $B_z$ profile.
%\subsection{Limitations of OpenGGCM}
%- Finite difference scheme (?) \\
%- Instabilities in the simulations \\
The second source of error is associated with the sensitivity of OpenGGCM to the first $B_z$ value provided as initial condition. The occurrence of rapidly and unnaturally varying features in the modeled Dst profile midway through the main phase of the storms and continuing all the way up to the recovery phase suggested the commencement of the accumulation of the instabilities in the simulations. So, we made a conscious strategic effort to choose the starting point such that $B_z$ was low but non-zero to avoid abrupt gradients in the simulation domain. Hence, the accuracy of Dst predictions of OpenGGCM is strongly correlated with the $B_z$ predictions of EUHFORIA. The final errors could be associated with the coupled ring current model with OpenGGCM which makes the Dst calculations. %Unfortunately, the OpenGGCM simulations are computationally expensive, i.e., they can take up to 24~hours to complete, which makes the model less suitable for operational space weather forecasting.

%%%%%%%%%%%%%%%%%%%%%%%%%%%%%%%%%%%%%%%%%%%%%%%%%%%%%%%%%%%%%
%-------------------------Section---------------------------%
%%%%%%%%%%%%%%%%%%%%%%%%%%%%%%%%%%%%%%%%%%%%%%%%%%%%%%%%%%%%%
\section{Summary and Discussion}
\label{sec:conclusion}
An overview of the methodologies adopted in this study is as follows:
\begin{itemize}
    \item We successfully demonstrate the coupling between the solar wind and CME evolution model, EUHFORIA, and the coupled magnetosphere-ionosphere-thermosphere model, OpenGGCM.
    \item The geomagnetic indices are derived by post-processing the output of this newly coupled model. The Dst index was computed using the ring current model, RCM, coupled to OpenGGCM. The A-indices were quantified by averaging the magnetic field perturbations in the ionosphere. 
    \item In addition, we have presented the validation of this coupling with the widely studied moderate-to-intense geomagnetic storms of July 2012 (Event 1) and September 2017 (Event 2). Two OpenGGCM simulations were performed for each event -- one employed the in situ observations (Event`n'-obs), and the other used the EUHFORIA output (Event`n'-euh) for initializing OpenGGCM. %The simulation initialized with in situ observations is used to benchmark the performance of OpenGGCM.
    \item We then assessed the predicted Dst profiles of Event`n'-obs and Event`n'-euh using the DTW technique to evaluate the whole range of Dst predictions, not just the minimum Dst point.  
\end{itemize}

In the case of Event 1 (CME on 12 July 2012) Dst prediction based on the in situ observations (Event1-obs) slightly outperformed the prediction made by EUHFORIA input (Event1-euh). The predicted AE indices, in both Event1-obs and Event1-euh, although qualitatively matched the order of magnitude of the observations during the period of the storm, a one-to-one correspondence between the extreme points was not attained. For Event 2 (CME on 4--6 September 2017), the {Dst predictions by OpenGGCM} with EUHFORIA input outperformed the one with in situ observations. In this two-step storm, the enhancements in the predicted AE indices were clustered around the two peaks as in the observations. AE indices were captured better for Event 2. We conclude that EUHFORIA-driven OpenGGCM simulations, although overestimating the Dst, are still in reasonable agreement in order of magnitude with the in situ driven OpenGGCM simulations. The A-indices predicted by the EUHFORIA-driven OpenGGCM simulations match qualitatively the overall periods and the magnitudes of the enhancements.

The two event studies were merely used to showcase the validation of the coupling. It is not possible to comment on why OpenGGCM predictions were better using EUHFORIA inputs over the actual L1 observations, in Event 2 but not in Event 1. %In addition to predicting geomagnetic indices with the coupled EUHFORIA and OpenGGCM model, 
Further studies need to be undertaken to understand the effect of different solar wind magnetic fields and flow speed profiles on the variability of the OpenGGCM predictions. Parametric studies with synthetic EUHFORIA simulation inputs can improve our understanding of the drivers of storms and substorms. %$P_{dyn}$ and especially $|B|$ are less relevant for both storms and substorms. We conclude that these directed information transfers constitute robust interrelationships between solar wind parameters and dynamic processes in the magnetosphere. 
EUHFORIA simulations can be performed as fast as 25 times the physical time of the process. With sufficient resources OpenGGCM can be run easily as fast as 10 times real time, making long-range predictions (typically several days) possible with the coupled models. This study is one of the efforts to couple MHD models of the heliosphere with CMEs and global magnetosphere to improve space weather forecasting, which lays the groundwork for future improvements. {The simulations discussed in this work show a potential to obtain forecasts 1--2 days in advance, with a similar accuracy as the current nowcasts.}

\acknowledgments
A.M.\ acknowledges support from the projects C14/19/089 (C1 project Internal Funds KU Leuven), G.0025.23N (WEAVE FWO-Vlaanderen), SIDC Data Exploitation (ESA Prodex-12), and Belspo project B2/191/P1/SWiM. E.S.\ research was supported by an appointment to the NASA Postdoctoral Program at the NASA Goddard Space Flight Center, administered by Oak Ridge Associated Universities under contract with NASA. Computations were performed on Marvin, a Cray CS500 supercomputer at UNH supported by the NSF MRI program under grant AGS-1919310. The research at UNH was supported by the Air Force Office of Scientific Research (grant no. FA9550-18-1-0483) and NASA (grant no. 80NSSC18K1220).

\section*{Open Data Statement}
The EUHFORIA (ver 2.0, \citeA{Pomoell2018}) simulations were performed on the Vlaams Supercomputer Centrum (VSC, \url{http://www.vscentrum.be}), Belgium. They can be reproduced by outside users on \url{https://www.euhforiaonline.com/} by creating an account through VSC. The input parameters of the simulations are provided in Section~\ref{sec:validation} of the paper. The OpenGGCM (v5.0.ccmc, \citeA{Raeder1998,Cramer2017}) simulations were carried out on Marvin cluster at the University of New Hampshire. The same simulations can be performed on the openly accessible Community Coordinated Modeling Center (CCMC, \url{https://ccmc.gsfc.nasa.gov/}). Both VSC and CCMC require free user subscription. The setup and input files, output files and the plotting scripts are available at \citeA{anweshaM2023}(\url{https://zenodo.org/doi/10.5281/zenodo.10404880}).
%% ------------------------------------------------------------------------ %%
%% References and Citations

%%%%%%%%%%%%%%%%%%%%%%%%%%%%%%%%%%%%%%%%%%%%%%%
%
% \bibliography{<name of your .bib file>} don't specify the file extension
%
% don't specify bibliographystyle

% In the References section, cite the data/software described in the Availability Statement (this includes primary and processed data used for your research). For details on data/software citation as well as examples, see the Data & Software Citation section of the Data & Software for Authors guidance
% https://www.agu.org/Publish-with-AGU/Publish/Author-Resources/Data-and-Software-for-Authors#citation

%%%%%%%%%%%%%%%%%%%%%%%%%%%%%%%%%%%%%%%%%%%%%%%

\bibliography{agusample.bib}

\begin{thebibliography}{}

\bibitem [\protect \citeauthoryear {%
{Akasofu}%
}{%
{Akasofu}%
}{%
{\protect \APACyear {1968}}%
}]{%
Akasofu1968}
\APACinsertmetastar {%
Akasofu1968}%
\begin{APACrefauthors}%
{Akasofu}, S\BPBI I.%
\end{APACrefauthors}%
\unskip\
\newblock
\APACrefYear{1968}.
\newblock
\APACrefbtitle {{Polar and Magnetosphere Substorms}} {{Polar and Magnetosphere
  Substorms}}\ (\BVOL~11).
\newblock
\begin{APACrefDOI} \doi{10.1007/978-94-010-3461-6} \end{APACrefDOI}
\PrintBackRefs{\CurrentBib}

\bibitem [\protect \citeauthoryear {%
{Akasofu}%
}{%
{Akasofu}%
}{%
{\protect \APACyear {2020}}%
}]{%
Akasofu2020}
\APACinsertmetastar {%
Akasofu2020}%
\begin{APACrefauthors}%
{Akasofu}, S\BHBI I.%
\end{APACrefauthors}%
\unskip\
\newblock
\APACrefYearMonthDay{2020}{{\APACmonth{12}}}{}.
\newblock
{\BBOQ}\APACrefatitle {{Relationship between geomagnetic storms and
  auroral/Magnetospheric substorms: Early studies}} {{Relationship between
  geomagnetic storms and auroral/Magnetospheric substorms: Early
  studies}}.{\BBCQ}
\newblock
\APACjournalVolNumPages{Frontiers in Astronomy and Space Sciences}{7}{}{101}.
\newblock
\begin{APACrefDOI} \doi{10.3389/fspas.2020.604755} \end{APACrefDOI}
\PrintBackRefs{\CurrentBib}

\bibitem [\protect \citeauthoryear {%
{Akasofu}%
}{%
{Akasofu}%
}{%
{\protect \APACyear {2021}}%
}]{%
Akasofu2021}
\APACinsertmetastar {%
Akasofu2021}%
\begin{APACrefauthors}%
{Akasofu}, S\BHBI I.%
\end{APACrefauthors}%
\unskip\
\newblock
\APACrefYearMonthDay{2021}{{\APACmonth{01}}}{}.
\newblock
{\BBOQ}\APACrefatitle {{A Review of Studies of Geomagnetic Storms and
  Auroral/Magnetospheric Substorms based on the Electric Current Approach}} {{A
  Review of Studies of Geomagnetic Storms and Auroral/Magnetospheric Substorms
  based on the Electric Current Approach}}.{\BBCQ}
\newblock
\APACjournalVolNumPages{Frontiers in Astronomy and Space Sciences}{7}{}{100}.
\newblock
\begin{APACrefDOI} \doi{10.3389/fspas.2020.604750} \end{APACrefDOI}
\PrintBackRefs{\CurrentBib}

\bibitem [\protect \citeauthoryear {%
{Akasofu}%
\ \BBA {} {Chapman}%
}{%
{Akasofu}%
\ \BBA {} {Chapman}%
}{%
{\protect \APACyear {1963}}%
}]{%
Akasofu1963}
\APACinsertmetastar {%
Akasofu1963}%
\begin{APACrefauthors}%
{Akasofu}, S\BPBI I.%
\BCBT {}\ \BBA {} {Chapman}, S.%
\end{APACrefauthors}%
\unskip\
\newblock
\APACrefYearMonthDay{1963}{{\APACmonth{01}}}{}.
\newblock
{\BBOQ}\APACrefatitle {{The Development of the Main Phase of Magnetic Storms}}
  {{The Development of the Main Phase of Magnetic Storms}}.{\BBCQ}
\newblock
\APACjournalVolNumPages{Journal of Geophysical Research (Space
  Physics)}{68}{1}{125-129}.
\newblock
\begin{APACrefDOI} \doi{10.1029/JZ068i001p00125} \end{APACrefDOI}
\PrintBackRefs{\CurrentBib}

\bibitem [\protect \citeauthoryear {%
{Alberti}%
\ \protect \BOthers {.}}{%
{Alberti}%
\ \protect \BOthers {.}}{%
{\protect \APACyear {2022}}%
}]{%
Alberti2022}
\APACinsertmetastar {%
Alberti2022}%
\begin{APACrefauthors}%
{Alberti}, T.%
, {Faranda}, D.%
, {Consolini}, G.%
, {De Michelis}, P.%
, {Donner}, R\BPBI V.%
\BCBL {}\ \BBA {} {Carbone}, V.%
\end{APACrefauthors}%
\unskip\
\newblock
\APACrefYearMonthDay{2022}{{\APACmonth{04}}}{}.
\newblock
{\BBOQ}\APACrefatitle {{Concurrent Effects between Geomagnetic Storms and
  Magnetospheric Substorms}} {{Concurrent Effects between Geomagnetic Storms
  and Magnetospheric Substorms}}.{\BBCQ}
\newblock
\APACjournalVolNumPages{Universe}{8}{4}{226}.
\newblock
\begin{APACrefDOI} \doi{10.3390/universe8040226} \end{APACrefDOI}
\PrintBackRefs{\CurrentBib}

\bibitem [\protect \citeauthoryear {%
{Arge}%
, {Luhmann}%
, {Odstrcil}%
, {Schrijver}%
\BCBL {}\ \BBA {} {Li}%
}{%
{Arge}%
\ \protect \BOthers {.}}{%
{\protect \APACyear {2004}}%
}]{%
arge2004}
\APACinsertmetastar {%
arge2004}%
\begin{APACrefauthors}%
{Arge}, C\BPBI N.%
, {Luhmann}, J\BPBI G.%
, {Odstrcil}, D.%
, {Schrijver}, C\BPBI J.%
\BCBL {}\ \BBA {} {Li}, Y.%
\end{APACrefauthors}%
\unskip\
\newblock
\APACrefYearMonthDay{2004}{{\APACmonth{10}}}{}.
\newblock
{\BBOQ}\APACrefatitle {{Stream structure and coronal sources of the solar wind
  during the May 12th, 1997 CME}} {{Stream structure and coronal sources of the
  solar wind during the May 12th, 1997 CME}}.{\BBCQ}
\newblock
\APACjournalVolNumPages{Journal of Atmospheric and Solar-Terrestrial
  Physics}{66}{15-16}{1295-1309}.
\newblock
\begin{APACrefDOI} \doi{10.1016/j.jastp.2004.03.018} \end{APACrefDOI}
\PrintBackRefs{\CurrentBib}

\bibitem [\protect \citeauthoryear {%
{Asvestari}%
\ \protect \BOthers {.}}{%
{Asvestari}%
\ \protect \BOthers {.}}{%
{\protect \APACyear {2019}}%
}]{%
Asvestari2019}
\APACinsertmetastar {%
Asvestari2019}%
\begin{APACrefauthors}%
{Asvestari}, E.%
, {Heinemann}, S\BPBI G.%
, {Temmer}, M.%
, {Pomoell}, J.%
, {Kilpua}, E.%
, {Magdalenic}, J.%
\BCBL {}\ \BBA {} {Poedts}, S.%
\end{APACrefauthors}%
\unskip\
\newblock
\APACrefYearMonthDay{2019}{{\APACmonth{11}}}{}.
\newblock
{\BBOQ}\APACrefatitle {{Reconstructing Coronal Hole Areas With EUHFORIA and
  Adapted WSA Model: Optimizing the Model Parameters}} {{Reconstructing Coronal
  Hole Areas With EUHFORIA and Adapted WSA Model: Optimizing the Model
  Parameters}}.{\BBCQ}
\newblock
\APACjournalVolNumPages{Journal of Geophysical Research (Space
  Physics)}{124}{11}{8280-8297}.
\newblock
\begin{APACrefDOI} \doi{10.1029/2019JA027173} \end{APACrefDOI}
\PrintBackRefs{\CurrentBib}

\bibitem [\protect \citeauthoryear {%
{Barnard}%
\ \BBA {} {Owens}%
}{%
{Barnard}%
\ \BBA {} {Owens}%
}{%
{\protect \APACyear {2022}}%
}]{%
Barnard2022}
\APACinsertmetastar {%
Barnard2022}%
\begin{APACrefauthors}%
{Barnard}, L.%
\BCBT {}\ \BBA {} {Owens}, M.%
\end{APACrefauthors}%
\unskip\
\newblock
\APACrefYearMonthDay{2022}{{\APACmonth{10}}}{}.
\newblock
{\BBOQ}\APACrefatitle {{HUXt{\textemdash}An open source, computationally
  efficient reduced-physics solar wind model, written in Python}}
  {{HUXt{\textemdash}An open source, computationally efficient reduced-physics
  solar wind model, written in Python}}.{\BBCQ}
\newblock
\APACjournalVolNumPages{Frontiers in Physics}{10}{}{1005621}.
\newblock
\begin{APACrefDOI} \doi{10.3389/fphy.2022.1005621} \end{APACrefDOI}
\PrintBackRefs{\CurrentBib}

\bibitem [\protect \citeauthoryear {%
{Berger}%
\ \protect \BOthers {.}}{%
{Berger}%
\ \protect \BOthers {.}}{%
{\protect \APACyear {2018}}%
}]{%
Berger2018}
\APACinsertmetastar {%
Berger2018}%
\begin{APACrefauthors}%
{Berger}, T.%
, {Matthi{\"a}}, D.%
, {Burmeister}, S.%
, {Rios}, R.%
, {Lee}, K.%
, {Semones}, E.%
\BDBL {}{Zeitlin}, C.%
\end{APACrefauthors}%
\unskip\
\newblock
\APACrefYearMonthDay{2018}{{\APACmonth{09}}}{}.
\newblock
{\BBOQ}\APACrefatitle {{The Solar Particle Event on 10 September 2017 as
  observed onboard the International Space Station (ISS)}} {{The Solar Particle
  Event on 10 September 2017 as observed onboard the International Space
  Station (ISS)}}.{\BBCQ}
\newblock
\APACjournalVolNumPages{Space Weather}{16}{9}{1173-1189}.
\newblock
\begin{APACrefDOI} \doi{10.1029/2018SW001920} \end{APACrefDOI}
\PrintBackRefs{\CurrentBib}

\bibitem [\protect \citeauthoryear {%
{Bergin}%
, {Chapman}%
\BCBL {}\ \BBA {} {Gjerloev}%
}{%
{Bergin}%
\ \protect \BOthers {.}}{%
{\protect \APACyear {2020}}%
}]{%
Bergin2020}
\APACinsertmetastar {%
Bergin2020}%
\begin{APACrefauthors}%
{Bergin}, A.%
, {Chapman}, S\BPBI C.%
\BCBL {}\ \BBA {} {Gjerloev}, J\BPBI W.%
\end{APACrefauthors}%
\unskip\
\newblock
\APACrefYearMonthDay{2020}{{\APACmonth{05}}}{}.
\newblock
{\BBOQ}\APACrefatitle {{AE, D$_{ST}$, and Their SuperMAG Counterparts: The
  Effect of Improved Spatial Resolution in Geomagnetic Indices}} {{AE,
  D$_{ST}$, and Their SuperMAG Counterparts: The Effect of Improved Spatial
  Resolution in Geomagnetic Indices}}.{\BBCQ}
\newblock
\APACjournalVolNumPages{Journal of Geophysical Research (Space
  Physics)}{125}{5}{e27828}.
\newblock
\begin{APACrefDOI} \doi{10.1029/2020JA027828} \end{APACrefDOI}
\PrintBackRefs{\CurrentBib}

\bibitem [\protect \citeauthoryear {%
{Boteler}%
}{%
{Boteler}%
}{%
{\protect \APACyear {2001}}%
}]{%
Boteler2001}
\APACinsertmetastar {%
Boteler2001}%
\begin{APACrefauthors}%
{Boteler}, D\BPBI H.%
\end{APACrefauthors}%
\unskip\
\newblock
\APACrefYearMonthDay{2001}{{\APACmonth{01}}}{}.
\newblock
{\BBOQ}\APACrefatitle {{Space weather effects on power systems}} {{Space
  weather effects on power systems}}.{\BBCQ}
\newblock
\APACjournalVolNumPages{Washington DC American Geophysical Union Geophysical
  Monograph Series}{125}{}{347-352}.
\newblock
\begin{APACrefDOI} \doi{10.1029/GM125p0347} \end{APACrefDOI}
\PrintBackRefs{\CurrentBib}

\bibitem [\protect \citeauthoryear {%
Bothmer%
\ \BBA {} Daglis%
}{%
Bothmer%
\ \BBA {} Daglis%
}{%
{\protect \APACyear {2007}}%
}]{%
Bothmer2007}
\APACinsertmetastar {%
Bothmer2007}%
\begin{APACrefauthors}%
Bothmer, V.%
\BCBT {}\ \BBA {} Daglis, I\BPBI A.%
\end{APACrefauthors}%
\unskip\
\newblock
\APACrefYear{2007}.
\newblock
\APACrefbtitle {Space weather: physics and effects} {Space weather: physics and
  effects}.
\newblock
\APACaddressPublisher{}{Springer Science \& Business Media}.
\PrintBackRefs{\CurrentBib}

\bibitem [\protect \citeauthoryear {%
{Burton}%
, {McPherron}%
\BCBL {}\ \BBA {} {Russell}%
}{%
{Burton}%
\ \protect \BOthers {.}}{%
{\protect \APACyear {1975}}%
}]{%
Burton1975}
\APACinsertmetastar {%
Burton1975}%
\begin{APACrefauthors}%
{Burton}, R\BPBI K.%
, {McPherron}, R\BPBI L.%
\BCBL {}\ \BBA {} {Russell}, C\BPBI T.%
\end{APACrefauthors}%
\unskip\
\newblock
\APACrefYearMonthDay{1975}{{\APACmonth{11}}}{}.
\newblock
{\BBOQ}\APACrefatitle {{An empirical relationship between interplanetary
  conditions and Dst}} {{An empirical relationship between interplanetary
  conditions and Dst}}.{\BBCQ}
\newblock
\APACjournalVolNumPages{Journal of Geophysical Research (Space
  Physics)}{80}{31}{4204}.
\newblock
\begin{APACrefDOI} \doi{10.1029/JA080i031p04204} \end{APACrefDOI}
\PrintBackRefs{\CurrentBib}

\bibitem [\protect \citeauthoryear {%
{Buzulukova}%
\ \BBA {} {Tsurutani}%
}{%
{Buzulukova}%
\ \BBA {} {Tsurutani}%
}{%
{\protect \APACyear {2022}}%
}]{%
Buzulukova2022}
\APACinsertmetastar {%
Buzulukova2022}%
\begin{APACrefauthors}%
{Buzulukova}, N.%
\BCBT {}\ \BBA {} {Tsurutani}, B.%
\end{APACrefauthors}%
\unskip\
\newblock
\APACrefYearMonthDay{2022}{{\APACmonth{12}}}{}.
\newblock
{\BBOQ}\APACrefatitle {{Space Weather: From Solar Origins to Risks and Hazards
  Evolving in Time}} {{Space Weather: From Solar Origins to Risks and Hazards
  Evolving in Time}}.{\BBCQ}
\newblock
\APACjournalVolNumPages{Frontiers in Astronomy and Space Sciences}{9}{}{429}.
\newblock
\begin{APACrefDOI} \doi{10.3389/fspas.2022.1017103} \end{APACrefDOI}
\PrintBackRefs{\CurrentBib}

\bibitem [\protect \citeauthoryear {%
{Chapman}%
}{%
{Chapman}%
}{%
{\protect \APACyear {1918}}%
}]{%
Chapman1918}
\APACinsertmetastar {%
Chapman1918}%
\begin{APACrefauthors}%
{Chapman}, S.%
\end{APACrefauthors}%
\unskip\
\newblock
\APACrefYearMonthDay{1918}{{\APACmonth{10}}}{}.
\newblock
{\BBOQ}\APACrefatitle {{An Outline of a Theory of Magnetic Storms}} {{An
  Outline of a Theory of Magnetic Storms}}.{\BBCQ}
\newblock
\APACjournalVolNumPages{Proceedings of the Royal Society of London Series
  A}{95}{666}{61-83}.
\newblock
\begin{APACrefDOI} \doi{10.1098/rspa.1918.0049} \end{APACrefDOI}
\PrintBackRefs{\CurrentBib}

\bibitem [\protect \citeauthoryear {%
{Chapman}%
\ \BBA {} {Ferraro}%
}{%
{Chapman}%
\ \BBA {} {Ferraro}%
}{%
{\protect \APACyear {1931}}%
}]{%
Chapman1931}
\APACinsertmetastar {%
Chapman1931}%
\begin{APACrefauthors}%
{Chapman}, S.%
\BCBT {}\ \BBA {} {Ferraro}, V\BPBI C\BPBI A.%
\end{APACrefauthors}%
\unskip\
\newblock
\APACrefYearMonthDay{1931}{{\APACmonth{01}}}{}.
\newblock
{\BBOQ}\APACrefatitle {{A new theory of magnetic storms}} {{A new theory of
  magnetic storms}}.{\BBCQ}
\newblock
\APACjournalVolNumPages{Terrestrial Magnetism and Atmospheric Electricity
  (Journal of Geophysical Research)}{36}{2}{77}.
\newblock
\begin{APACrefDOI} \doi{10.1029/TE036i002p00077} \end{APACrefDOI}
\PrintBackRefs{\CurrentBib}

\bibitem [\protect \citeauthoryear {%
{Chiu}%
\ \protect \BOthers {.}}{%
{Chiu}%
\ \protect \BOthers {.}}{%
{\protect \APACyear {1998}}%
}]{%
Chiu1998}
\APACinsertmetastar {%
Chiu1998}%
\begin{APACrefauthors}%
{Chiu}, M\BPBI C.%
, {von-Mehlem}, U\BPBI I.%
, {Willey}, C\BPBI E.%
, {Betenbaugh}, T\BPBI M.%
, {Maynard}, J\BPBI J.%
, {Krein}, J\BPBI A.%
\BDBL {}{Rodberg}, E\BPBI H.%
\end{APACrefauthors}%
\unskip\
\newblock
\APACrefYearMonthDay{1998}{{\APACmonth{07}}}{}.
\newblock
{\BBOQ}\APACrefatitle {{ACE Spacecraft}} {{ACE Spacecraft}}.{\BBCQ}
\newblock
\APACjournalVolNumPages{Space Science Reviews}{86}{}{257-284}.
\newblock
\begin{APACrefDOI} \doi{10.1023/A:1005002013459} \end{APACrefDOI}
\PrintBackRefs{\CurrentBib}

\bibitem [\protect \citeauthoryear {%
{Cohen}%
\ \BBA {} {Mewaldt}%
}{%
{Cohen}%
\ \BBA {} {Mewaldt}%
}{%
{\protect \APACyear {2018}}%
}]{%
Cohen2018}
\APACinsertmetastar {%
Cohen2018}%
\begin{APACrefauthors}%
{Cohen}, C\BPBI M\BPBI S.%
\BCBT {}\ \BBA {} {Mewaldt}, R\BPBI A.%
\end{APACrefauthors}%
\unskip\
\newblock
\APACrefYearMonthDay{2018}{{\APACmonth{10}}}{}.
\newblock
{\BBOQ}\APACrefatitle {{The Ground-Level Enhancement Event of September 2017
  and Other Large Solar Energetic Particle Events of Cycle 24}} {{The
  Ground-Level Enhancement Event of September 2017 and Other Large Solar
  Energetic Particle Events of Cycle 24}}.{\BBCQ}
\newblock
\APACjournalVolNumPages{Space Weather}{16}{10}{1616-1623}.
\newblock
\begin{APACrefDOI} \doi{10.1029/2018SW002006} \end{APACrefDOI}
\PrintBackRefs{\CurrentBib}

\bibitem [\protect \citeauthoryear {%
{Cramer}%
, {Raeder}%
, {Toffoletto}%
, {Gilson}%
\BCBL {}\ \BBA {} {Hu}%
}{%
{Cramer}%
\ \protect \BOthers {.}}{%
{\protect \APACyear {2017}}%
}]{%
Cramer2017}
\APACinsertmetastar {%
Cramer2017}%
\begin{APACrefauthors}%
{Cramer}, W\BPBI D.%
, {Raeder}, J.%
, {Toffoletto}, F\BPBI R.%
, {Gilson}, M.%
\BCBL {}\ \BBA {} {Hu}, B.%
\end{APACrefauthors}%
\unskip\
\newblock
\APACrefYearMonthDay{2017}{{\APACmonth{05}}}{}.
\newblock
{\BBOQ}\APACrefatitle {{Plasma sheet injections into the inner magnetosphere:
  Two-way coupled OpenGGCM-RCM model results}} {{Plasma sheet injections into
  the inner magnetosphere: Two-way coupled OpenGGCM-RCM model results}}.{\BBCQ}
\newblock
\APACjournalVolNumPages{Journal of Geophysical Research (Space
  Physics)}{122}{5}{5077-5091}.
\newblock
\begin{APACrefDOI} \doi{10.1002/2017JA024104} \end{APACrefDOI}
\PrintBackRefs{\CurrentBib}

\bibitem [\protect \citeauthoryear {%
{Davis}%
\ \BBA {} {Sugiura}%
}{%
{Davis}%
\ \BBA {} {Sugiura}%
}{%
{\protect \APACyear {1966}}%
}]{%
Davis1966}
\APACinsertmetastar {%
Davis1966}%
\begin{APACrefauthors}%
{Davis}, T\BPBI N.%
\BCBT {}\ \BBA {} {Sugiura}, M.%
\end{APACrefauthors}%
\unskip\
\newblock
\APACrefYearMonthDay{1966}{{\APACmonth{02}}}{}.
\newblock
{\BBOQ}\APACrefatitle {{Auroral electrojet activity index AE and its universal
  time variations}} {{Auroral electrojet activity index AE and its universal
  time variations}}.{\BBCQ}
\newblock
\APACjournalVolNumPages{Journal of Geophysical Research (Space
  Physics)}{71}{3}{785-801}.
\newblock
\begin{APACrefDOI} \doi{10.1029/JZ071i003p00785} \end{APACrefDOI}
\PrintBackRefs{\CurrentBib}

\bibitem [\protect \citeauthoryear {%
{DeJong}%
, {Bell}%
\BCBL {}\ \BBA {} {Ridley}%
}{%
{DeJong}%
\ \protect \BOthers {.}}{%
{\protect \APACyear {2018}}%
}]{%
dejong2018}
\APACinsertmetastar {%
dejong2018}%
\begin{APACrefauthors}%
{DeJong}, A\BPBI D.%
, {Bell}, J\BPBI M.%
\BCBL {}\ \BBA {} {Ridley}, A.%
\end{APACrefauthors}%
\unskip\
\newblock
\APACrefYearMonthDay{2018}{{\APACmonth{06}}}{}.
\newblock
{\BBOQ}\APACrefatitle {{Comparison of the Ionosphere During an SMC Initiating
  Substorm and an Isolated Substorm}} {{Comparison of the Ionosphere During an
  SMC Initiating Substorm and an Isolated Substorm}}.{\BBCQ}
\newblock
\APACjournalVolNumPages{Journal of Geophysical Research (Space
  Physics)}{123}{6}{4939-4951}.
\newblock
\begin{APACrefDOI} \doi{10.1029/2017JA025055} \end{APACrefDOI}
\PrintBackRefs{\CurrentBib}

\bibitem [\protect \citeauthoryear {%
{Dessler}%
\ \BBA {} {Parker}%
}{%
{Dessler}%
\ \BBA {} {Parker}%
}{%
{\protect \APACyear {1959}}%
}]{%
Dessler1959}
\APACinsertmetastar {%
Dessler1959}%
\begin{APACrefauthors}%
{Dessler}, A\BPBI J.%
\BCBT {}\ \BBA {} {Parker}, E\BPBI N.%
\end{APACrefauthors}%
\unskip\
\newblock
\APACrefYearMonthDay{1959}{{\APACmonth{12}}}{}.
\newblock
{\BBOQ}\APACrefatitle {{Hydromagnetic theory of geomagnetic storms}}
  {{Hydromagnetic theory of geomagnetic storms}}.{\BBCQ}
\newblock
\APACjournalVolNumPages{Journal of Geophysical Research (Space
  Physics)}{64}{12}{2239-2252}.
\newblock
\begin{APACrefDOI} \doi{10.1029/JZ064i012p02239} \end{APACrefDOI}
\PrintBackRefs{\CurrentBib}

\bibitem [\protect \citeauthoryear {%
{Dodson}%
\ \BBA {} {Hedeman}%
}{%
{Dodson}%
\ \BBA {} {Hedeman}%
}{%
{\protect \APACyear {1964}}%
}]{%
dodson1964}
\APACinsertmetastar {%
dodson1964}%
\begin{APACrefauthors}%
{Dodson}, H\BPBI W.%
\BCBT {}\ \BBA {} {Hedeman}, E\BPBI R.%
\end{APACrefauthors}%
\unskip\
\newblock
\APACrefYearMonthDay{1964}{May}{}.
\newblock
{\BBOQ}\APACrefatitle {{Problems of differentiation of flares with respect to
  geophysical effects}} {{Problems of differentiation of flares with respect to
  geophysical effects}}.{\BBCQ}
\newblock
\APACjournalVolNumPages{Planetary Space Science}{12}{5}{393-418}.
\newblock
\begin{APACrefDOI} \doi{10.1016/0032-0633(64)90034-0} \end{APACrefDOI}
\PrintBackRefs{\CurrentBib}

\bibitem [\protect \citeauthoryear {%
{Dungey}%
}{%
{Dungey}%
}{%
{\protect \APACyear {1961}}%
}]{%
Dungey1961}
\APACinsertmetastar {%
Dungey1961}%
\begin{APACrefauthors}%
{Dungey}, J\BPBI W.%
\end{APACrefauthors}%
\unskip\
\newblock
\APACrefYearMonthDay{1961}{{\APACmonth{01}}}{}.
\newblock
{\BBOQ}\APACrefatitle {{Interplanetary Magnetic Field and the Auroral Zones}}
  {{Interplanetary Magnetic Field and the Auroral Zones}}.{\BBCQ}
\newblock
\APACjournalVolNumPages{Physical Review Letters}{6}{2}{47-48}.
\newblock
\begin{APACrefDOI} \doi{10.1103/PhysRevLett.6.47} \end{APACrefDOI}
\PrintBackRefs{\CurrentBib}

\bibitem [\protect \citeauthoryear {%
{Evans}%
\ \BBA {} {Hawley}%
}{%
{Evans}%
\ \BBA {} {Hawley}%
}{%
{\protect \APACyear {1988}}%
}]{%
Evans1988}
\APACinsertmetastar {%
Evans1988}%
\begin{APACrefauthors}%
{Evans}, C\BPBI R.%
\BCBT {}\ \BBA {} {Hawley}, J\BPBI F.%
\end{APACrefauthors}%
\unskip\
\newblock
\APACrefYearMonthDay{1988}{{\APACmonth{09}}}{}.
\newblock
{\BBOQ}\APACrefatitle {{Simulation of Magnetohydrodynamic Flows: A Constrained
  Transport Model}} {{Simulation of Magnetohydrodynamic Flows: A Constrained
  Transport Model}}.{\BBCQ}
\newblock
\APACjournalVolNumPages{Astrophysical Journal}{332}{}{659}.
\newblock
\begin{APACrefDOI} \doi{10.1086/166684} \end{APACrefDOI}
\PrintBackRefs{\CurrentBib}

\bibitem [\protect \citeauthoryear {%
FREng%
}{%
FREng%
}{%
{\protect \APACyear {2013}}%
}]{%
FREng2007}
\APACinsertmetastar {%
FREng2007}%
\begin{APACrefauthors}%
FREng, P\BPBI C.%
\end{APACrefauthors}%
\unskip\
\newblock
\APACrefYear{2013}.
\newblock
\APACrefbtitle {Extreme space weather: impacts on engineered systems and
  infrastructure} {Extreme space weather: impacts on engineered systems and
  infrastructure}.
\newblock
\APACaddressPublisher{}{Royal Academy of Engineering}.
\PrintBackRefs{\CurrentBib}

\bibitem [\protect \citeauthoryear {%
{Fuller-Rowell}%
, {Codrescu}%
, {Risbeth}%
, {Moffett}%
\BCBL {}\ \BBA {} {Quegan}%
}{%
{Fuller-Rowell}%
\ \protect \BOthers {.}}{%
{\protect \APACyear {1996}}%
}]{%
Fuller-Rowell1996}
\APACinsertmetastar {%
Fuller-Rowell1996}%
\begin{APACrefauthors}%
{Fuller-Rowell}, T\BPBI J.%
, {Codrescu}, M\BPBI V.%
, {Risbeth}, H.%
, {Moffett}, R\BPBI J.%
\BCBL {}\ \BBA {} {Quegan}, S.%
\end{APACrefauthors}%
\unskip\
\newblock
\APACrefYearMonthDay{1996}{{\APACmonth{02}}}{}.
\newblock
{\BBOQ}\APACrefatitle {{On the seasonal response of the thermosphere and
  ionosphere to geomagnetic storms}} {{On the seasonal response of the
  thermosphere and ionosphere to geomagnetic storms}}.{\BBCQ}
\newblock
\APACjournalVolNumPages{Journal of Geophysical Research (Space
  Physics)}{101}{A2}{2343-2354}.
\newblock
\begin{APACrefDOI} \doi{10.1029/95JA01614} \end{APACrefDOI}
\PrintBackRefs{\CurrentBib}

\bibitem [\protect \citeauthoryear {%
{Fuller-Rowell}%
\ \BBA {} {Rees}%
}{%
{Fuller-Rowell}%
\ \BBA {} {Rees}%
}{%
{\protect \APACyear {1983}}%
}]{%
Fuller-Rowell1983}
\APACinsertmetastar {%
Fuller-Rowell1983}%
\begin{APACrefauthors}%
{Fuller-Rowell}, T\BPBI J.%
\BCBT {}\ \BBA {} {Rees}, D.%
\end{APACrefauthors}%
\unskip\
\newblock
\APACrefYearMonthDay{1983}{{\APACmonth{10}}}{}.
\newblock
{\BBOQ}\APACrefatitle {{Derivation of a conservation equation for mean
  molecular weight for a two-constituent gas within a three-dimensional,
  time-dependent model of the thermosphere}} {{Derivation of a conservation
  equation for mean molecular weight for a two-constituent gas within a
  three-dimensional, time-dependent model of the thermosphere}}.{\BBCQ}
\newblock
\APACjournalVolNumPages{Planetary Space Science}{31}{10}{1209-1222}.
\newblock
\begin{APACrefDOI} \doi{10.1016/0032-0633(83)90112-5} \end{APACrefDOI}
\PrintBackRefs{\CurrentBib}

\bibitem [\protect \citeauthoryear {%
{Gopalswamy}%
, {Akiyama}%
, {Yashiro}%
\BCBL {}\ \BBA {} {Xie}%
}{%
{Gopalswamy}%
\ \protect \BOthers {.}}{%
{\protect \APACyear {2018}}%
}]{%
Gopalswamy2018}
\APACinsertmetastar {%
Gopalswamy2018}%
\begin{APACrefauthors}%
{Gopalswamy}, N.%
, {Akiyama}, S.%
, {Yashiro}, S.%
\BCBL {}\ \BBA {} {Xie}, H.%
\end{APACrefauthors}%
\unskip\
\newblock
\APACrefYearMonthDay{2018}{{\APACmonth{11}}}{}.
\newblock
{\BBOQ}\APACrefatitle {{Coronal flux ropes and their interplanetary
  counterparts}} {{Coronal flux ropes and their interplanetary
  counterparts}}.{\BBCQ}
\newblock
\APACjournalVolNumPages{Journal of Atmospheric and Solar-Terrestrial
  Physics}{180}{}{35-45}.
\newblock
\begin{APACrefDOI} \doi{10.1016/j.jastp.2017.06.004} \end{APACrefDOI}
\PrintBackRefs{\CurrentBib}

\bibitem [\protect \citeauthoryear {%
G{\'o}recki%
\ \BBA {} {\L}uczak%
}{%
G{\'o}recki%
\ \BBA {} {\L}uczak%
}{%
{\protect \APACyear {2013}}%
}]{%
Gorecki2013}
\APACinsertmetastar {%
Gorecki2013}%
\begin{APACrefauthors}%
G{\'o}recki, T.%
\BCBT {}\ \BBA {} {\L}uczak, M.%
\end{APACrefauthors}%
\unskip\
\newblock
\APACrefYearMonthDay{2013}{Mar}{01}.
\newblock
{\BBOQ}\APACrefatitle {Using derivatives in time series classification} {Using
  derivatives in time series classification}.{\BBCQ}
\newblock
\APACjournalVolNumPages{Data Mining and Knowledge Discovery}{26}{2}{310-331}.
\newblock
\begin{APACrefURL} \url{https://doi.org/10.1007/s10618-012-0251-4}
  \end{APACrefURL}
\newblock
\begin{APACrefDOI} \doi{10.1007/s10618-012-0251-4} \end{APACrefDOI}
\PrintBackRefs{\CurrentBib}

\bibitem [\protect \citeauthoryear {%
H.~{Hu}%
, {Liu}%
, {Wang}%
, {M{\"o}stl}%
\BCBL {}\ \BBA {} {Yang}%
}{%
H.~{Hu}%
\ \protect \BOthers {.}}{%
{\protect \APACyear {2016}}%
}]{%
Hu2016}
\APACinsertmetastar {%
Hu2016}%
\begin{APACrefauthors}%
{Hu}, H.%
, {Liu}, Y\BPBI D.%
, {Wang}, R.%
, {M{\"o}stl}, C.%
\BCBL {}\ \BBA {} {Yang}, Z.%
\end{APACrefauthors}%
\unskip\
\newblock
\APACrefYearMonthDay{2016}{{\APACmonth{10}}}{}.
\newblock
{\BBOQ}\APACrefatitle {{Sun-to-Earth Characteristics of the 2012 July 12
  Coronal Mass Ejection and Associated Geo-effectiveness}} {{Sun-to-Earth
  Characteristics of the 2012 July 12 Coronal Mass Ejection and Associated
  Geo-effectiveness}}.{\BBCQ}
\newblock
\APACjournalVolNumPages{Astrophysical Journal}{829}{2}{97}.
\newblock
\begin{APACrefDOI} \doi{10.3847/0004-637X/829/2/97} \end{APACrefDOI}
\PrintBackRefs{\CurrentBib}

\bibitem [\protect \citeauthoryear {%
Y\BPBI Q.~{Hu}%
, {Guo}%
\BCBL {}\ \BBA {} {Wang}%
}{%
Y\BPBI Q.~{Hu}%
\ \protect \BOthers {.}}{%
{\protect \APACyear {2007}}%
}]{%
Hu2007}
\APACinsertmetastar {%
Hu2007}%
\begin{APACrefauthors}%
{Hu}, Y\BPBI Q.%
, {Guo}, X\BPBI C.%
\BCBL {}\ \BBA {} {Wang}, C.%
\end{APACrefauthors}%
\unskip\
\newblock
\APACrefYearMonthDay{2007}{{\APACmonth{07}}}{}.
\newblock
{\BBOQ}\APACrefatitle {{On the ionospheric and reconnection potentials of the
  earth: Results from global MHD simulations}} {{On the ionospheric and
  reconnection potentials of the earth: Results from global MHD
  simulations}}.{\BBCQ}
\newblock
\APACjournalVolNumPages{Journal of Geophysical Research (Space
  Physics)}{112}{A7}{A07215}.
\newblock
\begin{APACrefDOI} \doi{10.1029/2006JA012145} \end{APACrefDOI}
\PrintBackRefs{\CurrentBib}

\bibitem [\protect \citeauthoryear {%
{Iwai}%
\ \protect \BOthers {.}}{%
{Iwai}%
\ \protect \BOthers {.}}{%
{\protect \APACyear {2021}}%
}]{%
Iwai2021}
\APACinsertmetastar {%
Iwai2021}%
\begin{APACrefauthors}%
{Iwai}, K.%
, {Shiota}, D.%
, {Tokumaru}, M.%
, {Fujiki}, K.%
, {Den}, M.%
\BCBL {}\ \BBA {} {Kubo}, Y.%
\end{APACrefauthors}%
\unskip\
\newblock
\APACrefYearMonthDay{2021}{{\APACmonth{12}}}{}.
\newblock
{\BBOQ}\APACrefatitle {{Validation of coronal mass ejection arrival-time
  forecasts by magnetohydrodynamic simulations based on interplanetary
  scintillation observations}} {{Validation of coronal mass ejection
  arrival-time forecasts by magnetohydrodynamic simulations based on
  interplanetary scintillation observations}}.{\BBCQ}
\newblock
\APACjournalVolNumPages{Earth, Planets and Space}{73}{1}{9}.
\newblock
\begin{APACrefDOI} \doi{10.1186/s40623-020-01345-5} \end{APACrefDOI}
\PrintBackRefs{\CurrentBib}

\bibitem [\protect \citeauthoryear {%
{Janhunen}%
, {Koskinen}%
\BCBL {}\ \BBA {} {Pulkinen}%
}{%
{Janhunen}%
\ \protect \BOthers {.}}{%
{\protect \APACyear {1996}}%
}]{%
Janhunen1996}
\APACinsertmetastar {%
Janhunen1996}%
\begin{APACrefauthors}%
{Janhunen}, P.%
, {Koskinen}, K\BPBI E\BPBI J.%
\BCBL {}\ \BBA {} {Pulkinen}, T\BPBI I.%
\end{APACrefauthors}%
\unskip\
\newblock
\APACrefYearMonthDay{1996}{{\APACmonth{10}}}{}.
\newblock
{\BBOQ}\APACrefatitle {{A new global ionosphere-magnetosphere coupling
  simulation utilizing locally varying time step}} {{A new global
  ionosphere-magnetosphere coupling simulation utilizing locally varying time
  step}}.{\BBCQ}
\newblock
\BIn{} E\BPBI J.~{Rolfe}\ \BBA {} B.~{Kaldeich}\ (\BEDS), \APACrefbtitle
  {International Conference on Substorms} {International conference on
  substorms}\ (\BVOL~389, \BPG~205).
\PrintBackRefs{\CurrentBib}

\bibitem [\protect \citeauthoryear {%
{Kataoka}%
, {Sato}%
, {Miyake}%
, {Shiota}%
\BCBL {}\ \BBA {} {Kubo}%
}{%
{Kataoka}%
\ \protect \BOthers {.}}{%
{\protect \APACyear {2018}}%
}]{%
Kataoka2018}
\APACinsertmetastar {%
Kataoka2018}%
\begin{APACrefauthors}%
{Kataoka}, R.%
, {Sato}, T.%
, {Miyake}, S.%
, {Shiota}, D.%
\BCBL {}\ \BBA {} {Kubo}, Y.%
\end{APACrefauthors}%
\unskip\
\newblock
\APACrefYearMonthDay{2018}{{\APACmonth{07}}}{}.
\newblock
{\BBOQ}\APACrefatitle {{Radiation Dose Nowcast for the Ground Level Enhancement
  on 10-11 September 2017}} {{Radiation Dose Nowcast for the Ground Level
  Enhancement on 10-11 September 2017}}.{\BBCQ}
\newblock
\APACjournalVolNumPages{Space Weather}{16}{7}{917-923}.
\newblock
\begin{APACrefDOI} \doi{10.1029/2018SW001874} \end{APACrefDOI}
\PrintBackRefs{\CurrentBib}

\bibitem [\protect \citeauthoryear {%
{Keesee}%
\ \protect \BOthers {.}}{%
{Keesee}%
\ \protect \BOthers {.}}{%
{\protect \APACyear {2020}}%
}]{%
Keesee2020}
\APACinsertmetastar {%
Keesee2020}%
\begin{APACrefauthors}%
{Keesee}, A\BPBI M.%
, {Pinto}, V.%
, {Coughlan}, M.%
, {Lennox}, C.%
, {Mahmud}, M\BPBI S.%
\BCBL {}\ \BBA {} {Connor}, H\BPBI K.%
\end{APACrefauthors}%
\unskip\
\newblock
\APACrefYearMonthDay{2020}{{\APACmonth{10}}}{}.
\newblock
{\BBOQ}\APACrefatitle {{Comparison of deep learning techniques to model
  connections between solar wind and ground magnetic perturbations}}
  {{Comparison of deep learning techniques to model connections between solar
  wind and ground magnetic perturbations}}.{\BBCQ}
\newblock
\APACjournalVolNumPages{Frontiers in Astronomy and Space Sciences}{7}{}{72}.
\newblock
\begin{APACrefDOI} \doi{10.3389/fspas.2020.550874} \end{APACrefDOI}
\PrintBackRefs{\CurrentBib}

\bibitem [\protect \citeauthoryear {%
{Keogh}%
\ \BBA {} {Pazzani}%
}{%
{Keogh}%
\ \BBA {} {Pazzani}%
}{%
{\protect \APACyear {2001}}%
}]{%
Keogh2001}
\APACinsertmetastar {%
Keogh2001}%
\begin{APACrefauthors}%
{Keogh}, E\BPBI J.%
\BCBT {}\ \BBA {} {Pazzani}, M\BPBI J.%
\end{APACrefauthors}%
\unskip\
\newblock
\APACrefYearMonthDay{2001}{}{}.
\newblock
{\BBOQ}\APACrefatitle {{}} {{}}.{\BBCQ}
\newblock
\APACjournalVolNumPages{SIAM Int. Conf. on Data Mining (SDM) (Philadelphia, PA:
  Society for Industrial and Applied Mathematics)}{}{}{}.
\PrintBackRefs{\CurrentBib}

\bibitem [\protect \citeauthoryear {%
Krausmann%
, Andersson%
, Gibbs%
\BCBL {}\ \BBA {} Murtagh%
}{%
Krausmann%
\ \protect \BOthers {.}}{%
{\protect \APACyear {2016}}%
}]{%
krausmann2016}
\APACinsertmetastar {%
krausmann2016}%
\begin{APACrefauthors}%
Krausmann, E.%
, Andersson, E.%
, Gibbs, M.%
\BCBL {}\ \BBA {} Murtagh, W.%
\end{APACrefauthors}%
\unskip\
\newblock
\APACrefYearMonthDay{2016}{}{}.
\newblock
\APACrefbtitle {Space weather \& Critical Infrastructures: Findings and
  Outlook} {Space weather \& critical infrastructures: Findings and outlook}\
  \APACbVolEdTR {}{Anticipation and foresight\ \BNUMS\ LB-NA-28237-EN-N
  (online),LB-NA-28237-EN-E (ePub)}.
\newblock
\APACaddressInstitution{Luxembourg (Luxembourg)}{}.
\newblock
\begin{APACrefDOI} \doi{10.2788/152877 (online),10.2788/54866 (ePub)}
  \end{APACrefDOI}
\PrintBackRefs{\CurrentBib}

\bibitem [\protect \citeauthoryear {%
{Laperre}%
, {Amaya}%
\BCBL {}\ \BBA {} {Lapenta}%
}{%
{Laperre}%
\ \protect \BOthers {.}}{%
{\protect \APACyear {2020}}%
}]{%
Laperre2020}
\APACinsertmetastar {%
Laperre2020}%
\begin{APACrefauthors}%
{Laperre}, B.%
, {Amaya}, J.%
\BCBL {}\ \BBA {} {Lapenta}, G.%
\end{APACrefauthors}%
\unskip\
\newblock
\APACrefYearMonthDay{2020}{{\APACmonth{07}}}{}.
\newblock
{\BBOQ}\APACrefatitle {{Dynamic Time Warping as a New Evaluation for Dst
  Forecast with Machine Learning}} {{Dynamic Time Warping as a New Evaluation
  for Dst Forecast with Machine Learning}}.{\BBCQ}
\newblock
\APACjournalVolNumPages{Frontiers in Astronomy and Space Sciences}{7}{}{39}.
\newblock
\begin{APACrefDOI} \doi{10.3389/fspas.2020.00039} \end{APACrefDOI}
\PrintBackRefs{\CurrentBib}

\bibitem [\protect \citeauthoryear {%
{Luhmann}%
\ \protect \BOthers {.}}{%
{Luhmann}%
\ \protect \BOthers {.}}{%
{\protect \APACyear {2004}}%
}]{%
Luhmann2004}
\APACinsertmetastar {%
Luhmann2004}%
\begin{APACrefauthors}%
{Luhmann}, J\BPBI G.%
, {Solomon}, S\BPBI C.%
, {Linker}, J\BPBI A.%
, {Lyon}, J\BPBI G.%
, {Mikic}, Z.%
, {Odstrcil}, D.%
\BDBL {}{Wiltberger}, M.%
\end{APACrefauthors}%
\unskip\
\newblock
\APACrefYearMonthDay{2004}{{\APACmonth{10}}}{}.
\newblock
{\BBOQ}\APACrefatitle {{Coupled model simulation of a Sun-to-Earth space
  weather event}} {{Coupled model simulation of a Sun-to-Earth space weather
  event}}.{\BBCQ}
\newblock
\APACjournalVolNumPages{Journal of Atmospheric and Solar-Terrestrial
  Physics}{66}{15-16}{1243-1256}.
\newblock
\begin{APACrefDOI} \doi{10.1016/j.jastp.2004.04.005} \end{APACrefDOI}
\PrintBackRefs{\CurrentBib}

\bibitem [\protect \citeauthoryear {%
{Maharana}%
\ \protect \BOthers {.}}{%
{Maharana}%
\ \protect \BOthers {.}}{%
{\protect \APACyear {2022}}%
}]{%
Maharana2022}
\APACinsertmetastar {%
Maharana2022}%
\begin{APACrefauthors}%
{Maharana}, A.%
, {Isavnin}, A.%
, {Scolini}, C.%
, {Wijsen}, N.%
, {Rodriguez}, L.%
, {Mierla}, M.%
\BDBL {}{Poedts}, S.%
\end{APACrefauthors}%
\unskip\
\newblock
\APACrefYearMonthDay{2022}{{\APACmonth{09}}}{}.
\newblock
{\BBOQ}\APACrefatitle {{Implementation and validation of the FRi3D flux rope
  model in EUHFORIA}} {{Implementation and validation of the FRi3D flux rope
  model in EUHFORIA}}.{\BBCQ}
\newblock
\APACjournalVolNumPages{Advances in Space Research}{70}{6}{1641-1662}.
\newblock
\begin{APACrefDOI} \doi{10.1016/j.asr.2022.05.056} \end{APACrefDOI}
\PrintBackRefs{\CurrentBib}

\bibitem [\protect \citeauthoryear {%
{Newell}%
\ \BBA {} {Gjerloev}%
}{%
{Newell}%
\ \BBA {} {Gjerloev}%
}{%
{\protect \APACyear {2011}}%
}]{%
Newell2011}
\APACinsertmetastar {%
Newell2011}%
\begin{APACrefauthors}%
{Newell}, P\BPBI T.%
\BCBT {}\ \BBA {} {Gjerloev}, J\BPBI W.%
\end{APACrefauthors}%
\unskip\
\newblock
\APACrefYearMonthDay{2011}{{\APACmonth{12}}}{}.
\newblock
{\BBOQ}\APACrefatitle {{Evaluation of SuperMAG auroral electrojet indices as
  indicators of substorms and auroral power}} {{Evaluation of SuperMAG auroral
  electrojet indices as indicators of substorms and auroral power}}.{\BBCQ}
\newblock
\APACjournalVolNumPages{Journal of Geophysical Research (Space
  Physics)}{116}{A12}{A12211}.
\newblock
\begin{APACrefDOI} \doi{10.1029/2011JA016779} \end{APACrefDOI}
\PrintBackRefs{\CurrentBib}

\bibitem [\protect \citeauthoryear {%
{O'Brien}%
\ \BBA {} {McPherron}%
}{%
{O'Brien}%
\ \BBA {} {McPherron}%
}{%
{\protect \APACyear {2000}}%
}]{%
Obrien2000a}
\APACinsertmetastar {%
Obrien2000a}%
\begin{APACrefauthors}%
{O'Brien}, T\BPBI P.%
\BCBT {}\ \BBA {} {McPherron}, R\BPBI L.%
\end{APACrefauthors}%
\unskip\
\newblock
\APACrefYearMonthDay{2000}{{\APACmonth{04}}}{}.
\newblock
{\BBOQ}\APACrefatitle {{An empirical phase space analysis of ring current
  dynamics: Solar wind control of injection and decay}} {{An empirical phase
  space analysis of ring current dynamics: Solar wind control of injection and
  decay}}.{\BBCQ}
\newblock
\APACjournalVolNumPages{Journal of Geophysical Research (Space
  Physics)}{105}{A4}{7707-7720}.
\newblock
\begin{APACrefDOI} \doi{10.1029/1998JA000437} \end{APACrefDOI}
\PrintBackRefs{\CurrentBib}

\bibitem [\protect \citeauthoryear {%
{Odstrcil}%
}{%
{Odstrcil}%
}{%
{\protect \APACyear {2003}}%
}]{%
Odstrcil2003}
\APACinsertmetastar {%
Odstrcil2003}%
\begin{APACrefauthors}%
{Odstrcil}, D.%
\end{APACrefauthors}%
\unskip\
\newblock
\APACrefYearMonthDay{2003}{{\APACmonth{08}}}{}.
\newblock
{\BBOQ}\APACrefatitle {{Modeling 3-D solar wind structure}} {{Modeling 3-D
  solar wind structure}}.{\BBCQ}
\newblock
\APACjournalVolNumPages{Advances in Space Research}{32}{4}{497-506}.
\newblock
\begin{APACrefDOI} \doi{10.1016/S0273-1177(03)00332-6} \end{APACrefDOI}
\PrintBackRefs{\CurrentBib}

\bibitem [\protect \citeauthoryear {%
{Ogilvie}%
\ \BBA {} {Desch}%
}{%
{Ogilvie}%
\ \BBA {} {Desch}%
}{%
{\protect \APACyear {1997}}%
}]{%
Ogilvie1997}
\APACinsertmetastar {%
Ogilvie1997}%
\begin{APACrefauthors}%
{Ogilvie}, K\BPBI W.%
\BCBT {}\ \BBA {} {Desch}, M\BPBI D.%
\end{APACrefauthors}%
\unskip\
\newblock
\APACrefYearMonthDay{1997}{{\APACmonth{01}}}{}.
\newblock
{\BBOQ}\APACrefatitle {{The wind spacecraft and its early scientific results}}
  {{The wind spacecraft and its early scientific results}}.{\BBCQ}
\newblock
\APACjournalVolNumPages{Advances in Space Research}{20}{4-5}{559-568}.
\newblock
\begin{APACrefDOI} \doi{10.1016/S0273-1177(97)00439-0} \end{APACrefDOI}
\PrintBackRefs{\CurrentBib}

\bibitem [\protect \citeauthoryear {%
Oughton%
\ \protect \BOthers {.}}{%
Oughton%
\ \protect \BOthers {.}}{%
{\protect \APACyear {2019}}%
}]{%
Oughton2019}
\APACinsertmetastar {%
Oughton2019}%
\begin{APACrefauthors}%
Oughton, E\BPBI J.%
, Hapgood, M.%
, Richardson, G\BPBI S.%
, Beggan, C\BPBI D.%
, Thomson, A\BPBI W\BPBI P.%
, Gibbs, M.%
\BDBL {}Horne, R\BPBI B.%
\end{APACrefauthors}%
\unskip\
\newblock
\APACrefYearMonthDay{2019}{{\APACmonth{05}}}{}.
\newblock
{\BBOQ}\APACrefatitle {A risk assessment framework for the socioeconomic
  impacts of electricity transmission infrastructure failure due to space
  weather: An application to the United Kingdom} {A risk assessment framework
  for the socioeconomic impacts of electricity transmission infrastructure
  failure due to space weather: An application to the united kingdom}.{\BBCQ}
\newblock
\APACjournalVolNumPages{Risk Anal.}{39}{5}{1022--1043}.
\PrintBackRefs{\CurrentBib}

\bibitem [\protect \citeauthoryear {%
Palacios%
, Guerrero%
, Cid%
, Saiz%
\BCBL {}\ \BBA {} Cerrato%
}{%
Palacios%
\ \protect \BOthers {.}}{%
{\protect \APACyear {2018}}%
}]{%
Palacios2018}
\APACinsertmetastar {%
Palacios2018}%
\begin{APACrefauthors}%
Palacios, J.%
, Guerrero, A.%
, Cid, C.%
, Saiz, E.%
\BCBL {}\ \BBA {} Cerrato, Y.%
\end{APACrefauthors}%
\unskip\
\newblock
\APACrefYearMonthDay{2018}{}{}.
\newblock
{\BBOQ}\APACrefatitle {Defining scale thresholds for geomagnetic storms through
  statistics} {Defining scale thresholds for geomagnetic storms through
  statistics}.{\BBCQ}
\newblock
\APACjournalVolNumPages{Natural Hazards and Earth System Sciences
  Discussions}{2018}{}{1--17}.
\newblock
\begin{APACrefURL} \url{https://nhess.copernicus.org/preprints/nhess-2018-92/}
  \end{APACrefURL}
\newblock
\begin{APACrefDOI} \doi{10.5194/nhess-2018-92} \end{APACrefDOI}
\PrintBackRefs{\CurrentBib}

\bibitem [\protect \citeauthoryear {%
{Pilipenko}%
}{%
{Pilipenko}%
}{%
{\protect \APACyear {2021}}%
}]{%
Pilipenko2021}
\APACinsertmetastar {%
Pilipenko2021}%
\begin{APACrefauthors}%
{Pilipenko}, V.%
\end{APACrefauthors}%
\unskip\
\newblock
\APACrefYearMonthDay{2021}{{\APACmonth{09}}}{}.
\newblock
{\BBOQ}\APACrefatitle {{Space weather impact on ground-based technological
  systems}} {{Space weather impact on ground-based technological
  systems}}.{\BBCQ}
\newblock
\APACjournalVolNumPages{Solar-Terrestrial Physics}{7}{3}{68-104}.
\newblock
\begin{APACrefDOI} \doi{10.12737/stp-73202106} \end{APACrefDOI}
\PrintBackRefs{\CurrentBib}

\bibitem [\protect \citeauthoryear {%
{Poedts}%
}{%
{Poedts}%
}{%
{\protect \APACyear {2019}}%
}]{%
Poedts2019}
\APACinsertmetastar {%
Poedts2019}%
\begin{APACrefauthors}%
{Poedts}, S.%
\end{APACrefauthors}%
\unskip\
\newblock
\APACrefYearMonthDay{2019}{{\APACmonth{01}}}{}.
\newblock
{\BBOQ}\APACrefatitle {{Forecasting space weather with EUHFORIA in the virtual
  space weather modeling centre}} {{Forecasting space weather with EUHFORIA in
  the virtual space weather modeling centre}}.{\BBCQ}
\newblock
\APACjournalVolNumPages{Plasma Physics and Controlled Fusion}{61}{1}{014011}.
\newblock
\begin{APACrefDOI} \doi{10.1088/1361-6587/aae048} \end{APACrefDOI}
\PrintBackRefs{\CurrentBib}

\bibitem [\protect \citeauthoryear {%
{Pomoell}%
\ \BBA {} {Poedts}%
}{%
{Pomoell}%
\ \BBA {} {Poedts}%
}{%
{\protect \APACyear {2018}}%
}]{%
Pomoell2018}
\APACinsertmetastar {%
Pomoell2018}%
\begin{APACrefauthors}%
{Pomoell}, J.%
\BCBT {}\ \BBA {} {Poedts}, S.%
\end{APACrefauthors}%
\unskip\
\newblock
\APACrefYearMonthDay{2018}{{\APACmonth{06}}}{}.
\newblock
{\BBOQ}\APACrefatitle {{EUHFORIA: European heliospheric forecasting information
  asset}} {{EUHFORIA: European heliospheric forecasting information
  asset}}.{\BBCQ}
\newblock
\APACjournalVolNumPages{Journal of Space Weather and Space Climate}{8}{}{A35}.
\newblock
\begin{APACrefDOI} \doi{10.1051/swsc/2018020} \end{APACrefDOI}
\PrintBackRefs{\CurrentBib}

\bibitem [\protect \citeauthoryear {%
{Powell}%
, {Roe}%
, {Linde}%
, {Gombosi}%
\BCBL {}\ \BBA {} {De Zeeuw}%
}{%
{Powell}%
\ \protect \BOthers {.}}{%
{\protect \APACyear {1999}}%
}]{%
Powell1999}
\APACinsertmetastar {%
Powell1999}%
\begin{APACrefauthors}%
{Powell}, K\BPBI G.%
, {Roe}, P\BPBI L.%
, {Linde}, T\BPBI J.%
, {Gombosi}, T\BPBI I.%
\BCBL {}\ \BBA {} {De Zeeuw}, D\BPBI L.%
\end{APACrefauthors}%
\unskip\
\newblock
\APACrefYearMonthDay{1999}{{\APACmonth{09}}}{}.
\newblock
{\BBOQ}\APACrefatitle {{A Solution-Adaptive Upwind Scheme for Ideal
  Magnetohydrodynamics}} {{A Solution-Adaptive Upwind Scheme for Ideal
  Magnetohydrodynamics}}.{\BBCQ}
\newblock
\APACjournalVolNumPages{Journal of Computational Physics}{154}{2}{284-309}.
\newblock
\begin{APACrefDOI} \doi{10.1006/jcph.1999.6299} \end{APACrefDOI}
\PrintBackRefs{\CurrentBib}

\bibitem [\protect \citeauthoryear {%
{Raeder}%
, {Berchem}%
\BCBL {}\ \BBA {} {Ashour-Abdalla}%
}{%
{Raeder}%
\ \protect \BOthers {.}}{%
{\protect \APACyear {1998}}%
}]{%
Raeder1998}
\APACinsertmetastar {%
Raeder1998}%
\begin{APACrefauthors}%
{Raeder}, J.%
, {Berchem}, J.%
\BCBL {}\ \BBA {} {Ashour-Abdalla}, M.%
\end{APACrefauthors}%
\unskip\
\newblock
\APACrefYearMonthDay{1998}{{\APACmonth{07}}}{}.
\newblock
{\BBOQ}\APACrefatitle {{The Geospace Environment Modeling Grand Challenge:
  Results from a Global Geospace Circulation Model}} {{The Geospace Environment
  Modeling Grand Challenge: Results from a Global Geospace Circulation
  Model}}.{\BBCQ}
\newblock
\APACjournalVolNumPages{Journal of Geophysical Research (Space
  Physics)}{103}{A7}{14787-14798}.
\newblock
\begin{APACrefDOI} \doi{10.1029/98JA00014} \end{APACrefDOI}
\PrintBackRefs{\CurrentBib}

\bibitem [\protect \citeauthoryear {%
{Raeder}%
, {Wang}%
\BCBL {}\ \BBA {} {Fuller-Rowell}%
}{%
{Raeder}%
, {Wang}%
\BCBL {}\ \BBA {} {Fuller-Rowell}%
}{%
{\protect \APACyear {2001}}%
}]{%
Raeder2001a}
\APACinsertmetastar {%
Raeder2001a}%
\begin{APACrefauthors}%
{Raeder}, J.%
, {Wang}, Y.%
\BCBL {}\ \BBA {} {Fuller-Rowell}, T\BPBI J.%
\end{APACrefauthors}%
\unskip\
\newblock
\APACrefYearMonthDay{2001}{{\APACmonth{01}}}{}.
\newblock
{\BBOQ}\APACrefatitle {{Geomagnetic storm simulation with a coupled
  magnetosphere-ionosphere-thermosphere model}} {{Geomagnetic storm simulation
  with a coupled magnetosphere-ionosphere-thermosphere model}}.{\BBCQ}
\newblock
\APACjournalVolNumPages{Washington DC American Geophysical Union Geophysical
  Monograph Series}{125}{}{377-384}.
\newblock
\begin{APACrefDOI} \doi{10.1029/GM125p0377} \end{APACrefDOI}
\PrintBackRefs{\CurrentBib}

\bibitem [\protect \citeauthoryear {%
{Raeder}%
, {Wang}%
, {Fuller-Rowell}%
\BCBL {}\ \BBA {} {Singer}%
}{%
{Raeder}%
, {Wang}%
, {Fuller-Rowell}%
\BCBL {}\ \BBA {} {Singer}%
}{%
{\protect \APACyear {2001}}%
}]{%
Raeder2001b}
\APACinsertmetastar {%
Raeder2001b}%
\begin{APACrefauthors}%
{Raeder}, J.%
, {Wang}, Y\BPBI L.%
, {Fuller-Rowell}, T\BPBI J.%
\BCBL {}\ \BBA {} {Singer}, H\BPBI J.%
\end{APACrefauthors}%
\unskip\
\newblock
\APACrefYearMonthDay{2001}{{\APACmonth{12}}}{}.
\newblock
{\BBOQ}\APACrefatitle {{Global Simulation of Magnetospheric Space Weather
  Effects of the Bastille Day Storm}} {{Global Simulation of Magnetospheric
  Space Weather Effects of the Bastille Day Storm}}.{\BBCQ}
\newblock
\APACjournalVolNumPages{Solar Physics}{204}{}{323-337}.
\newblock
\begin{APACrefDOI} \doi{10.1023/A:1014228230714} \end{APACrefDOI}
\PrintBackRefs{\CurrentBib}

\bibitem [\protect \citeauthoryear {%
{Rast{\"a}tter}%
\ \protect \BOthers {.}}{%
{Rast{\"a}tter}%
\ \protect \BOthers {.}}{%
{\protect \APACyear {2013}}%
}]{%
Rastatter2013}
\APACinsertmetastar {%
Rastatter2013}%
\begin{APACrefauthors}%
{Rast{\"a}tter}, L.%
, {Kuznetsova}, M\BPBI M.%
, {Glocer}, A.%
, {Welling}, D.%
, {Meng}, X.%
, {Raeder}, J.%
\BDBL {}{Gannon}, J.%
\end{APACrefauthors}%
\unskip\
\newblock
\APACrefYearMonthDay{2013}{{\APACmonth{04}}}{}.
\newblock
{\BBOQ}\APACrefatitle {{Geospace environment modeling 2008-2009 challenge:
  D$_{st}$ index}} {{Geospace environment modeling 2008-2009 challenge:
  D$_{st}$ index}}.{\BBCQ}
\newblock
\APACjournalVolNumPages{Space Weather}{11}{4}{187-205}.
\newblock
\begin{APACrefDOI} \doi{10.1002/swe.20036} \end{APACrefDOI}
\PrintBackRefs{\CurrentBib}

\bibitem [\protect \citeauthoryear {%
{Redmon}%
, {Seaton}%
, {Steenburgh}%
, {He}%
\BCBL {}\ \BBA {} {Rodriguez}%
}{%
{Redmon}%
\ \protect \BOthers {.}}{%
{\protect \APACyear {2018}}%
}]{%
Redmon2018}
\APACinsertmetastar {%
Redmon2018}%
\begin{APACrefauthors}%
{Redmon}, R\BPBI J.%
, {Seaton}, D\BPBI B.%
, {Steenburgh}, R.%
, {He}, J.%
\BCBL {}\ \BBA {} {Rodriguez}, J\BPBI V.%
\end{APACrefauthors}%
\unskip\
\newblock
\APACrefYearMonthDay{2018}{{\APACmonth{09}}}{}.
\newblock
{\BBOQ}\APACrefatitle {{September 2017's Geoeffective Space Weather and Impacts
  to Caribbean Radio Communications During Hurricane Response}} {{September
  2017's Geoeffective Space Weather and Impacts to Caribbean Radio
  Communications During Hurricane Response}}.{\BBCQ}
\newblock
\APACjournalVolNumPages{Space Weather}{16}{9}{1190-1201}.
\newblock
\begin{APACrefDOI} \doi{10.1029/2018SW001897} \end{APACrefDOI}
\PrintBackRefs{\CurrentBib}

\bibitem [\protect \citeauthoryear {%
{Samara}%
\ \protect \BOthers {.}}{%
{Samara}%
\ \protect \BOthers {.}}{%
{\protect \APACyear {2022}}%
}]{%
Samara2022}
\APACinsertmetastar {%
Samara2022}%
\begin{APACrefauthors}%
{Samara}, E.%
, {Laperre}, B.%
, {Kieokaew}, R.%
, {Temmer}, M.%
, {Verbeke}, C.%
, {Rodriguez}, L.%
\BDBL {}{Poedts}, S.%
\end{APACrefauthors}%
\unskip\
\newblock
\APACrefYearMonthDay{2022}{{\APACmonth{03}}}{}.
\newblock
{\BBOQ}\APACrefatitle {{Dynamic Time Warping as a Means of Assessing Solar Wind
  Time Series}} {{Dynamic Time Warping as a Means of Assessing Solar Wind Time
  Series}}.{\BBCQ}
\newblock
\APACjournalVolNumPages{Astrophysical Journal}{927}{2}{187}.
\newblock
\begin{APACrefDOI} \doi{10.3847/1538-4357/ac4af6} \end{APACrefDOI}
\PrintBackRefs{\CurrentBib}

\bibitem [\protect \citeauthoryear {%
{Scolini}%
\ \protect \BOthers {.}}{%
{Scolini}%
\ \protect \BOthers {.}}{%
{\protect \APACyear {2020}}%
}]{%
Scolini2020}
\APACinsertmetastar {%
Scolini2020}%
\begin{APACrefauthors}%
{Scolini}, C.%
, {Chan{\'e}}, E.%
, {Temmer}, M.%
, {Kilpua}, E\BPBI K\BPBI J.%
, {Dissauer}, K.%
, {Veronig}, A\BPBI M.%
\BDBL {}{Poedts}, S.%
\end{APACrefauthors}%
\unskip\
\newblock
\APACrefYearMonthDay{2020}{{\APACmonth{02}}}{}.
\newblock
{\BBOQ}\APACrefatitle {{CME-CME Interactions as Sources of CME
  Geoeffectiveness: The Formation of the Complex Ejecta and Intense Geomagnetic
  Storm in 2017 Early September}} {{CME-CME Interactions as Sources of CME
  Geoeffectiveness: The Formation of the Complex Ejecta and Intense Geomagnetic
  Storm in 2017 Early September}}.{\BBCQ}
\newblock
\APACjournalVolNumPages{Astrophysical Journal, Supplement}{247}{1}{21}.
\newblock
\begin{APACrefDOI} \doi{10.3847/1538-4365/ab6216} \end{APACrefDOI}
\PrintBackRefs{\CurrentBib}

\bibitem [\protect \citeauthoryear {%
{Scolini}%
, {Rodriguez}%
, {Mierla}%
, {Pomoell}%
\BCBL {}\ \BBA {} {Poedts}%
}{%
{Scolini}%
\ \protect \BOthers {.}}{%
{\protect \APACyear {2019}}%
}]{%
Scolini2019}
\APACinsertmetastar {%
Scolini2019}%
\begin{APACrefauthors}%
{Scolini}, C.%
, {Rodriguez}, L.%
, {Mierla}, M.%
, {Pomoell}, J.%
\BCBL {}\ \BBA {} {Poedts}, S.%
\end{APACrefauthors}%
\unskip\
\newblock
\APACrefYearMonthDay{2019}{{\APACmonth{06}}}{}.
\newblock
{\BBOQ}\APACrefatitle {{Observation-based modelling of magnetised coronal mass
  ejections with EUHFORIA}} {{Observation-based modelling of magnetised coronal
  mass ejections with EUHFORIA}}.{\BBCQ}
\newblock
\APACjournalVolNumPages{Astronomy and Astrophysics}{626}{}{A122}.
\newblock
\begin{APACrefDOI} \doi{10.1051/0004-6361/201935053} \end{APACrefDOI}
\PrintBackRefs{\CurrentBib}

\bibitem [\protect \citeauthoryear {%
{Scolini}%
\ \protect \BOthers {.}}{%
{Scolini}%
\ \protect \BOthers {.}}{%
{\protect \APACyear {2018}}%
}]{%
Scolini2018}
\APACinsertmetastar {%
Scolini2018}%
\begin{APACrefauthors}%
{Scolini}, C.%
, {Verbeke}, C.%
, {Poedts}, S.%
, {Chan{\'e}}, E.%
, {Pomoell}, J.%
\BCBL {}\ \BBA {} {Zuccarello}, F\BPBI P.%
\end{APACrefauthors}%
\unskip\
\newblock
\APACrefYearMonthDay{2018}{{\APACmonth{06}}}{}.
\newblock
{\BBOQ}\APACrefatitle {{Effect of the Initial Shape of Coronal Mass Ejections
  on 3-D MHD Simulations and Geoeffectiveness Predictions}} {{Effect of the
  Initial Shape of Coronal Mass Ejections on 3-D MHD Simulations and
  Geoeffectiveness Predictions}}.{\BBCQ}
\newblock
\APACjournalVolNumPages{Space Weather}{16}{6}{754-771}.
\newblock
\begin{APACrefDOI} \doi{10.1029/2018SW001806} \end{APACrefDOI}
\PrintBackRefs{\CurrentBib}

\bibitem [\protect \citeauthoryear {%
{Shiota}%
\ \BBA {} {Kataoka}%
}{%
{Shiota}%
\ \BBA {} {Kataoka}%
}{%
{\protect \APACyear {2016}}%
}]{%
Shiota2016}
\APACinsertmetastar {%
Shiota2016}%
\begin{APACrefauthors}%
{Shiota}, D.%
\BCBT {}\ \BBA {} {Kataoka}, R.%
\end{APACrefauthors}%
\unskip\
\newblock
\APACrefYearMonthDay{2016}{{\APACmonth{02}}}{}.
\newblock
{\BBOQ}\APACrefatitle {{Magnetohydrodynamic simulation of interplanetary
  propagation of multiple coronal mass ejections with internal magnetic flux
  rope (SUSANOO-CME)}} {{Magnetohydrodynamic simulation of interplanetary
  propagation of multiple coronal mass ejections with internal magnetic flux
  rope (SUSANOO-CME)}}.{\BBCQ}
\newblock
\APACjournalVolNumPages{Space Weather}{14}{2}{56-75}.
\newblock
\begin{APACrefDOI} \doi{10.1002/2015SW001308} \end{APACrefDOI}
\PrintBackRefs{\CurrentBib}

\bibitem [\protect \citeauthoryear {%
{Skone}%
\ \BBA {} {de Jong}%
}{%
{Skone}%
\ \BBA {} {de Jong}%
}{%
{\protect \APACyear {2000}}%
}]{%
Skone2000}
\APACinsertmetastar {%
Skone2000}%
\begin{APACrefauthors}%
{Skone}, S.%
\BCBT {}\ \BBA {} {de Jong}, M.%
\end{APACrefauthors}%
\unskip\
\newblock
\APACrefYearMonthDay{2000}{{\APACmonth{11}}}{}.
\newblock
{\BBOQ}\APACrefatitle {{The impact of geomagnetic substorms on GPS receiver
  performance}} {{The impact of geomagnetic substorms on GPS receiver
  performance}}.{\BBCQ}
\newblock
\APACjournalVolNumPages{Earth, Planets and Space}{52}{}{1067-1071}.
\newblock
\begin{APACrefDOI} \doi{10.1186/BF03352332} \end{APACrefDOI}
\PrintBackRefs{\CurrentBib}

\bibitem [\protect \citeauthoryear {%
{Smith}%
\ \protect \BOthers {.}}{%
{Smith}%
\ \protect \BOthers {.}}{%
{\protect \APACyear {2021}}%
}]{%
Smith2021}
\APACinsertmetastar {%
Smith2021}%
\begin{APACrefauthors}%
{Smith}, A\BPBI W.%
, {Forsyth}, C.%
, {Rae}, I\BPBI J.%
, {Garton}, T\BPBI M.%
, {Bloch}, T.%
, {Jackman}, C\BPBI M.%
\BCBL {}\ \BBA {} {Bakrania}, M.%
\end{APACrefauthors}%
\unskip\
\newblock
\APACrefYearMonthDay{2021}{{\APACmonth{09}}}{}.
\newblock
{\BBOQ}\APACrefatitle {{Forecasting the Probability of Large Rates of Change of
  the Geomagnetic Field in the UK: Timescales, Horizons, and Thresholds}}
  {{Forecasting the Probability of Large Rates of Change of the Geomagnetic
  Field in the UK: Timescales, Horizons, and Thresholds}}.{\BBCQ}
\newblock
\APACjournalVolNumPages{Space Weather}{19}{9}{e2021SW002788}.
\newblock
\begin{APACrefDOI} \doi{10.1029/2021SW002788} \end{APACrefDOI}
\PrintBackRefs{\CurrentBib}

\bibitem [\protect \citeauthoryear {%
{Smith}%
\ \protect \BOthers {.}}{%
{Smith}%
\ \protect \BOthers {.}}{%
{\protect \APACyear {2022}}%
}]{%
Smith2022}
\APACinsertmetastar {%
Smith2022}%
\begin{APACrefauthors}%
{Smith}, A\BPBI W.%
, {Forsyth}, C.%
, {Rae}, I\BPBI J.%
, {Garton}, T\BPBI M.%
, {Jackman}, C\BPBI M.%
, {Bakrania}, M.%
\BDBL {}{Johnson}, J\BPBI M.%
\end{APACrefauthors}%
\unskip\
\newblock
\APACrefYearMonthDay{2022}{{\APACmonth{07}}}{}.
\newblock
{\BBOQ}\APACrefatitle {{On the Considerations of Using Near Real Time Data for
  Space Weather Hazard Forecasting}} {{On the Considerations of Using Near Real
  Time Data for Space Weather Hazard Forecasting}}.{\BBCQ}
\newblock
\APACjournalVolNumPages{Space Weather}{20}{7}{e2022SW003098}.
\newblock
\begin{APACrefDOI} \doi{10.1029/2022SW003098} \end{APACrefDOI}
\PrintBackRefs{\CurrentBib}

\bibitem [\protect \citeauthoryear {%
{Toffoletto}%
, {Sazykin}%
, {Spiro}%
\BCBL {}\ \BBA {} {Wolf}%
}{%
{Toffoletto}%
\ \protect \BOthers {.}}{%
{\protect \APACyear {2003}}%
}]{%
Toffoletto2003}
\APACinsertmetastar {%
Toffoletto2003}%
\begin{APACrefauthors}%
{Toffoletto}, F.%
, {Sazykin}, S.%
, {Spiro}, R.%
\BCBL {}\ \BBA {} {Wolf}, R.%
\end{APACrefauthors}%
\unskip\
\newblock
\APACrefYearMonthDay{2003}{{\APACmonth{04}}}{}.
\newblock
{\BBOQ}\APACrefatitle {{Inner magnetospheric modeling with the Rice Convection
  Model}} {{Inner magnetospheric modeling with the Rice Convection
  Model}}.{\BBCQ}
\newblock
\APACjournalVolNumPages{Space Science Reviews}{107}{1}{175-196}.
\newblock
\begin{APACrefDOI} \doi{10.1023/A:1025532008047} \end{APACrefDOI}
\PrintBackRefs{\CurrentBib}

\bibitem [\protect \citeauthoryear {%
{T{\'o}th}%
\ \protect \BOthers {.}}{%
{T{\'o}th}%
\ \protect \BOthers {.}}{%
{\protect \APACyear {2007}}%
}]{%
Toth2007}
\APACinsertmetastar {%
Toth2007}%
\begin{APACrefauthors}%
{T{\'o}th}, G.%
, {de Zeeuw}, D\BPBI L.%
, {Gombosi}, T\BPBI I.%
, {Manchester}, W\BPBI B.%
, {Ridley}, A\BPBI J.%
, {Sokolov}, I\BPBI V.%
\BCBL {}\ \BBA {} {Roussev}, I\BPBI I.%
\end{APACrefauthors}%
\unskip\
\newblock
\APACrefYearMonthDay{2007}{{\APACmonth{06}}}{}.
\newblock
{\BBOQ}\APACrefatitle {{Sun-to-thermosphere simulation of the 28-30 October
  2003 storm with the Space Weather Modeling Framework}} {{Sun-to-thermosphere
  simulation of the 28-30 October 2003 storm with the Space Weather Modeling
  Framework}}.{\BBCQ}
\newblock
\APACjournalVolNumPages{Space Weather}{5}{6}{06003}.
\newblock
\begin{APACrefDOI} \doi{10.1029/2006SW000272} \end{APACrefDOI}
\PrintBackRefs{\CurrentBib}

\bibitem [\protect \citeauthoryear {%
{Turner}%
\ \protect \BOthers {.}}{%
{Turner}%
\ \protect \BOthers {.}}{%
{\protect \APACyear {2001}}%
}]{%
Turner2001}
\APACinsertmetastar {%
Turner2001}%
\begin{APACrefauthors}%
{Turner}, N\BPBI E.%
, {Baker}, D\BPBI N.%
, {Pulkkinen}, T\BPBI I.%
, {Roeder}, J\BPBI L.%
, {Fennell}, J\BPBI F.%
\BCBL {}\ \BBA {} {Jordanova}, V\BPBI K.%
\end{APACrefauthors}%
\unskip\
\newblock
\APACrefYearMonthDay{2001}{{\APACmonth{09}}}{}.
\newblock
{\BBOQ}\APACrefatitle {{Energy content in the storm time ring current}}
  {{Energy content in the storm time ring current}}.{\BBCQ}
\newblock
\APACjournalVolNumPages{Journal of Geophysical Research (Space
  Physics)}{106}{A9}{19149-19156}.
\newblock
\begin{APACrefDOI} \doi{10.1029/2000JA003025} \end{APACrefDOI}
\PrintBackRefs{\CurrentBib}

\bibitem [\protect \citeauthoryear {%
{van der Holst}%
\ \protect \BOthers {.}}{%
{van der Holst}%
\ \protect \BOthers {.}}{%
{\protect \APACyear {2014}}%
}]{%
vanderholst2014}
\APACinsertmetastar {%
vanderholst2014}%
\begin{APACrefauthors}%
{van der Holst}, B.%
, {Sokolov}, I\BPBI V.%
, {Meng}, X.%
, {Jin}, M.%
, {Manchester}, I., W.~B.%
, {T{\'o}th}, G.%
\BCBL {}\ \BBA {} {Gombosi}, T\BPBI I.%
\end{APACrefauthors}%
\unskip\
\newblock
\APACrefYearMonthDay{2014}{{\APACmonth{02}}}{}.
\newblock
{\BBOQ}\APACrefatitle {{Alfv{\'e}n Wave Solar Model (AWSoM): Coronal Heating}}
  {{Alfv{\'e}n Wave Solar Model (AWSoM): Coronal Heating}}.{\BBCQ}
\newblock
\APACjournalVolNumPages{Astrophysical Journal}{782}{2}{81}.
\newblock
\begin{APACrefDOI} \doi{10.1088/0004-637X/782/2/81} \end{APACrefDOI}
\PrintBackRefs{\CurrentBib}

\bibitem [\protect \citeauthoryear {%
Verbeke%
\ \protect \BOthers {.}}{%
Verbeke%
\ \protect \BOthers {.}}{%
{\protect \APACyear {2022}}%
}]{%
VERBEKE2022}
\APACinsertmetastar {%
VERBEKE2022}%
\begin{APACrefauthors}%
Verbeke, C.%
, Mays, M\BPBI L.%
, Kay, C.%
, Riley, P.%
, Palmerio, E.%
, Dumbović, M.%
\BDBL {}Hinterreiter, J.%
\end{APACrefauthors}%
\unskip\
\newblock
\APACrefYearMonthDay{2022}{}{}.
\newblock
{\BBOQ}\APACrefatitle {Quantifying errors in 3D CME parameters derived from
  synthetic data using white-light reconstruction techniques} {Quantifying
  errors in 3d cme parameters derived from synthetic data using white-light
  reconstruction techniques}.{\BBCQ}
\newblock
\APACjournalVolNumPages{Advances in Space Research}{}{}{}.
\newblock
\begin{APACrefURL}
  \url{https://www.sciencedirect.com/science/article/pii/S027311772200792X}
  \end{APACrefURL}
\newblock
\begin{APACrefDOI} \doi{https://doi.org/10.1016/j.asr.2022.08.056}
  \end{APACrefDOI}
\PrintBackRefs{\CurrentBib}

\bibitem [\protect \citeauthoryear {%
{Verbeke}%
, {Pomoell}%
\BCBL {}\ \BBA {} {Poedts}%
}{%
{Verbeke}%
\ \protect \BOthers {.}}{%
{\protect \APACyear {2019}}%
}]{%
Verbeke2019}
\APACinsertmetastar {%
Verbeke2019}%
\begin{APACrefauthors}%
{Verbeke}, C.%
, {Pomoell}, J.%
\BCBL {}\ \BBA {} {Poedts}, S.%
\end{APACrefauthors}%
\unskip\
\newblock
\APACrefYearMonthDay{2019}{{\APACmonth{07}}}{}.
\newblock
{\BBOQ}\APACrefatitle {{The evolution of coronal mass ejections in the inner
  heliosphere: Implementing the spheromak model with EUHFORIA}} {{The evolution
  of coronal mass ejections in the inner heliosphere: Implementing the
  spheromak model with EUHFORIA}}.{\BBCQ}
\newblock
\APACjournalVolNumPages{Astronomy and Astrophysics}{627}{}{A111}.
\newblock
\begin{APACrefDOI} \doi{10.1051/0004-6361/201834702} \end{APACrefDOI}
\PrintBackRefs{\CurrentBib}

\bibitem [\protect \citeauthoryear {%
{Vr{\v{s}}nak}%
\ \protect \BOthers {.}}{%
{Vr{\v{s}}nak}%
\ \protect \BOthers {.}}{%
{\protect \APACyear {2013}}%
}]{%
vrsnak2013}
\APACinsertmetastar {%
vrsnak2013}%
\begin{APACrefauthors}%
{Vr{\v{s}}nak}, B.%
, {{\v{Z}}ic}, T.%
, {Vrbanec}, D.%
, {Temmer}, M.%
, {Rollett}, T.%
, {M{\"o}stl}, C.%
\BDBL {}{Shanmugaraju}, A.%
\end{APACrefauthors}%
\unskip\
\newblock
\APACrefYearMonthDay{2013}{{\APACmonth{07}}}{}.
\newblock
{\BBOQ}\APACrefatitle {{Propagation of Interplanetary Coronal Mass Ejections:
  The Drag-Based Model}} {{Propagation of Interplanetary Coronal Mass
  Ejections: The Drag-Based Model}}.{\BBCQ}
\newblock
\APACjournalVolNumPages{Solar Physics}{285}{1-2}{295-315}.
\newblock
\begin{APACrefDOI} \doi{10.1007/s11207-012-0035-4} \end{APACrefDOI}
\PrintBackRefs{\CurrentBib}

\bibitem [\protect \citeauthoryear {%
Wang%
\ \protect \BOthers {.}}{%
Wang%
\ \protect \BOthers {.}}{%
{\protect \APACyear {2013}}%
}]{%
Wang2013}
\APACinsertmetastar {%
Wang2013}%
\begin{APACrefauthors}%
Wang, C.%
, Guo, X.%
, Peng, Z.%
, Tang, B.%
, Sun, T.%
, Li, W.%
\BCBL {}\ \BBA {} Hu, Y.%
\end{APACrefauthors}%
\unskip\
\newblock
\APACrefYearMonthDay{2013}{Jul}{01}.
\newblock
{\BBOQ}\APACrefatitle {Magnetohydrodynamics (MHD) numerical simulations on the
  interaction of the solar wind with the magnetosphere: A review}
  {Magnetohydrodynamics (mhd) numerical simulations on the interaction of the
  solar wind with the magnetosphere: A review}.{\BBCQ}
\newblock
\APACjournalVolNumPages{Science China Earth Sciences}{56}{7}{1141-1157}.
\newblock
\begin{APACrefURL} \url{https://doi.org/10.1007/s11430-013-4608-3}
  \end{APACrefURL}
\newblock
\begin{APACrefDOI} \doi{10.1007/s11430-013-4608-3} \end{APACrefDOI}
\PrintBackRefs{\CurrentBib}

\bibitem [\protect \citeauthoryear {%
{Wanliss}%
\ \BBA {} {Showalter}%
}{%
{Wanliss}%
\ \BBA {} {Showalter}%
}{%
{\protect \APACyear {2006}}%
}]{%
Wanliss2006}
\APACinsertmetastar {%
Wanliss2006}%
\begin{APACrefauthors}%
{Wanliss}, J\BPBI A.%
\BCBT {}\ \BBA {} {Showalter}, K\BPBI M.%
\end{APACrefauthors}%
\unskip\
\newblock
\APACrefYearMonthDay{2006}{{\APACmonth{02}}}{}.
\newblock
{\BBOQ}\APACrefatitle {{High-resolution global storm index: Dst versus SYM-H}}
  {{High-resolution global storm index: Dst versus SYM-H}}.{\BBCQ}
\newblock
\APACjournalVolNumPages{Journal of Geophysical Research (Space
  Physics)}{111}{A2}{A02202}.
\newblock
\begin{APACrefDOI} \doi{10.1029/2005JA011034} \end{APACrefDOI}
\PrintBackRefs{\CurrentBib}

\bibitem [\protect \citeauthoryear {%
{Webb}%
\ \BBA {} {Nitta}%
}{%
{Webb}%
\ \BBA {} {Nitta}%
}{%
{\protect \APACyear {2017}}%
}]{%
Webb2017}
\APACinsertmetastar {%
Webb2017}%
\begin{APACrefauthors}%
{Webb}, D.%
\BCBT {}\ \BBA {} {Nitta}, N.%
\end{APACrefauthors}%
\unskip\
\newblock
\APACrefYearMonthDay{2017}{{\APACmonth{10}}}{}.
\newblock
{\BBOQ}\APACrefatitle {{Understanding Problem Forecasts of ISEST Campaign
  Flare-CME Events}} {{Understanding Problem Forecasts of ISEST Campaign
  Flare-CME Events}}.{\BBCQ}
\newblock
\APACjournalVolNumPages{Solar Physics}{292}{10}{142}.
\newblock
\begin{APACrefDOI} \doi{10.1007/s11207-017-1166-4} \end{APACrefDOI}
\PrintBackRefs{\CurrentBib}

\bibitem [\protect \citeauthoryear {%
{Wintoft}%
, {Wik}%
, {Matzka}%
\BCBL {}\ \BBA {} {Shprits}%
}{%
{Wintoft}%
\ \protect \BOthers {.}}{%
{\protect \APACyear {2017}}%
}]{%
Wintoft2017}
\APACinsertmetastar {%
Wintoft2017}%
\begin{APACrefauthors}%
{Wintoft}, P.%
, {Wik}, M.%
, {Matzka}, J.%
\BCBL {}\ \BBA {} {Shprits}, Y.%
\end{APACrefauthors}%
\unskip\
\newblock
\APACrefYearMonthDay{2017}{{\APACmonth{11}}}{}.
\newblock
{\BBOQ}\APACrefatitle {{Forecasting Kp from solar wind data: input parameter
  study using 3-hour averages and 3-hour range values}} {{Forecasting Kp from
  solar wind data: input parameter study using 3-hour averages and 3-hour range
  values}}.{\BBCQ}
\newblock
\APACjournalVolNumPages{Journal of Space Weather and Space Climate}{7}{}{A29}.
\newblock
\begin{APACrefDOI} \doi{10.1051/swsc/2017027} \end{APACrefDOI}
\PrintBackRefs{\CurrentBib}

\bibitem [\protect \citeauthoryear {%
{Yasyukevich}%
\ \protect \BOthers {.}}{%
{Yasyukevich}%
\ \protect \BOthers {.}}{%
{\protect \APACyear {2018}}%
}]{%
Yasyukevich2018}
\APACinsertmetastar {%
Yasyukevich2018}%
\begin{APACrefauthors}%
{Yasyukevich}, Y.%
, {Astafyeva}, E.%
, {Padokhin}, A.%
, {Ivanova}, V.%
, {Syrovatskii}, S.%
\BCBL {}\ \BBA {} {Podlesnyi}, A.%
\end{APACrefauthors}%
\unskip\
\newblock
\APACrefYearMonthDay{2018}{{\APACmonth{08}}}{}.
\newblock
{\BBOQ}\APACrefatitle {{The 6 September 2017 X-Class Solar Flares and Their
  Impacts on the Ionosphere, GNSS, and HF Radio Wave Propagation}} {{The 6
  September 2017 X-Class Solar Flares and Their Impacts on the Ionosphere,
  GNSS, and HF Radio Wave Propagation}}.{\BBCQ}
\newblock
\APACjournalVolNumPages{Space Weather}{16}{8}{1013-1027}.
\newblock
\begin{APACrefDOI} \doi{10.1029/2018SW001932} \end{APACrefDOI}
\PrintBackRefs{\CurrentBib}

\bibitem [\protect \citeauthoryear {%
Zenchenko%
\ \BBA {} Breus%
}{%
Zenchenko%
\ \BBA {} Breus%
}{%
{\protect \APACyear {2021}}%
}]{%
Zenchenko2021}
\APACinsertmetastar {%
Zenchenko2021}%
\begin{APACrefauthors}%
Zenchenko, T\BPBI A.%
\BCBT {}\ \BBA {} Breus, T\BPBI K.%
\end{APACrefauthors}%
\unskip\
\newblock
\APACrefYearMonthDay{2021}{}{}.
\newblock
{\BBOQ}\APACrefatitle {The Possible Effect of Space Weather Factors on Various
  Physiological Systems of the Human Organism} {The possible effect of space
  weather factors on various physiological systems of the human
  organism}.{\BBCQ}
\newblock
\APACjournalVolNumPages{Atmosphere}{12}{3}{}.
\newblock
\begin{APACrefURL} \url{https://www.mdpi.com/2073-4433/12/3/346}
  \end{APACrefURL}
\newblock
\begin{APACrefDOI} \doi{10.3390/atmos12030346} \end{APACrefDOI}
\PrintBackRefs{\CurrentBib}

\end{thebibliography}

%Reference citation instructions and examples:
%
% Please use ONLY \cite and \citeA for reference citations.
% \cite for parenthetical references
% ...as shown in recent studies (Simpson et al., 2019)
% \citeA for in-text citations
% ...Simpson et al. (2019) have shown...
%
%
%...as shown by \citeA{jskilby}.
%...as shown by \citeA{lewin76}, \citeA{carson86}, \citeA{bartoldy02}, and \citeA{rinaldi03}.
%...has been shown \cite{jskilbye}.
%...has been shown \cite{lewin76,carson86,bartoldy02,rinaldi03}.
%... \cite <i.e.>[]{lewin76,carson86,bartoldy02,rinaldi03}.
%...has been shown by \cite <e.g.,>[and others]{lewin76}.
%
% apacite uses < > for prenotes and [ ] for postnotes
% DO NOT use other cite commands (e.g., \citet, \citep, \citeyear, \citealp, etc.).
% \nocite is okay to use to add references from your Supporting Information
%

\end{document}